\documentclass[journal]{IEEEtran}  

\usepackage{multirow}
\usepackage{diagbox}
\usepackage{amssymb}
\usepackage{amsmath}
\usepackage{graphicx}
\usepackage{epstopdf}
\usepackage{epsfig}
\usepackage{cite}                
\usepackage{subfigure}
\usepackage{comment}
\usepackage{algorithm}
\usepackage{algpseudocode}
\usepackage{bm}
\usepackage{microtype}
\usepackage{url}
\usepackage{xcolor}
\usepackage{fancyhdr}
\usepackage{mdwmath}
\usepackage{mdwtab}
\usepackage[labelsep=period]{caption} 

\usepackage{amsthm}
\usepackage{lipsum}

\newtheorem{theorem}{Theorem}

\newtheorem{lemma}{Lemma}

\newtheorem{corollary}{Corollary}

\newtheorem{remark}{Remark}

\makeatletter
\def\ScaleIfNeeded{%
\ifdim\Gin@nat@width>\linewidth \linewidth \else \Gin@nat@width
\fi } \makeatother

\begin{document}

\title{NOMA Assisted Coordinated Beam Hopping for VLEO Satellite Communication Systems}

\author{Jiayang~Kang,~\IEEEmembership{Graduate Student Member,~IEEE},
        Xinwei~Yue,~\IEEEmembership{Senior Member,~IEEE},
        Xuyang~Zhang,~\\
        Tian~Li,~\IEEEmembership{Member,~IEEE},
        Peng~Yang,~\IEEEmembership{Member,~IEEE},
        Chunyu~Pan,~\IEEEmembership{Member,~IEEE},
        and~Shaoli~Kang,~\IEEEmembership{Member,~IEEE}
\thanks{J.~Kang, X.~Yue, and C.~Pan are with the College of Information and Communication Engineering, and with the Center for Target Cognition Information Processing Science and Technology, Beijing Information Science \& Technology University, Beijing 102206, China, and also with the Key Laboratory of Modern Measurement \& Control Technology, Ministry of Education, Beijing 102206, China (e-mail: \{jiayang.kang, xinwei.yue, chunyupan\}@bistu.edu.cn).}%
\thanks{X.~Zhang is with the School of Cyberspace Science and Technology, Beijing Institute of Technology, Beijing 100081, China (e-mail: 3220245313@bit.edu.cn).}%
\thanks{P.~Yang is with the School of Electronic and Information Engineering, and also with the State Key Laboratory of CNS/ATM, Beihang University, Beijing 100191, China (e-mail: yangp09@buaa.edu.cn).}%
\thanks{T.~Li is with the 54th Research Institute of China Electronics Technology Group Corporation, Shijiazhuang 050081, China (e-mail: t.li@ieee.org).}%
\thanks{S.~Kang is with the State Key Laboratory of Wireless Mobile Communications (CICT), Beijing 100191, China (e-mail: kangshaoli@cictmobile.com).}}

\maketitle

\begin{abstract}
This paper investigates a non-orthogonal multiple access assisted coordinated beam hopping (NOMA-CBH) scheme in multidimensional very low Earth orbit (VLEO) satellite communication systems, where multiple satellites employ CBH and coordinated beams jointly serve ground users and a dedicated ground station (GS). To characterize the VLEO-specific propagation environment, we develop a channel model that incorporates atmospheric attenuation, Doppler effects, and pointing perturbations. Building upon this architecture and channel model, a low-complexity decoupled (LCD) algorithm is designed to optimize beam scheduling and power allocation. We further derive closed-form outage probability expressions for the ground users and the GS, perform asymptotic analysis to characterize their diversity orders, and evaluate the delay-limited throughput. Simulation results demonstrate that: i) The NOMA-CBH with LCD algorithm achieves an enhanced system sum rate relative to orthogonal multiple access with coordinated beam hopping (OMA-CBH); ii) Across the considered heterogeneous traffic-demand operating points, the NOMA-CBH scheme achieves higher aggregate delivered traffic than the benchmark schemes; and iii) The outage performance of NOMA-CBH is better than that of OMA-CBH.
\end{abstract}


\section{Introduction}
With the development of forthcoming sixth-generation (6G) communication systems, space--air--ground integrated networks are envisioned as a key enabler for seamless global coverage and ubiquitous connectivity~\cite{xiaoyue2025zongshu,yang2025communications}. In this context, low Earth orbit (LEO) satellite systems have emerged as pivotal enablers of non-terrestrial communications~\cite{Azari2022leozongshu}. Nevertheless, conventional LEO systems operating at higher orbital altitudes generally experience longer propagation delays and less favorable link budgets, which can pose challenges for latency-sensitive and highly reliable 6G services~\cite{rehman2026vleo}. By contrast, very low Earth orbit (VLEO) satellites, typically deployed at altitudes below 450 km, markedly reduce propagation delay and enhance link quality, thereby emerging as a promising solution for 6G applications~\cite{mcDonald2024asta}. However, the very low orbital altitude introduces intensified atmospheric drag and, together with the rapid orbital motion and lower-altitude visibility geometry, results in rapid footprint variation and shorter service windows~\cite{crisp2021asta}. These characteristics render VLEO satellite systems highly dynamic, and the resulting uneven spatio-temporal traffic motivates the use of spatio-temporal beam scheduling, as previously studied in LEO satellite systems~\cite{leilei2024leoop}. However, extending these LEO resource-allocation and beam-scheduling formulations to VLEO requires explicit consideration of the rapid geometric variations, residual Doppler effects, atmospheric attenuation, and pointing perturbations associated with very low orbital altitudes~\cite{hrshawi2023losxindao}.

The attitude dynamics of VLEO satellites under extreme conditions were investigated in~\cite{zhang2025raodong}, where attitude-control and target-pointing challenges were shown to become increasingly pronounced at lower orbital altitudes. Furthermore, the authors of \cite{ferrato2025zhisun} investigated the aerodynamic effects of the residual atmosphere at ultra-low altitudes. Owing to the high orbital velocity of VLEO satellites, severe Doppler shifts were evaluated in~\cite{kim2025vleodoppler}, in which complex carrier synchronization and residual losses were systematically analyzed. To address the complex physical environment of VLEO systems, the authors of~\cite{shaun2020gaodu} characterized satellite aerodynamics and highlighted the related thermosphere interactions. Focusing on environmental torques, the authors in~\cite{lyu2026error} indicated that satellite attitude perturbations can induce antenna misalignment and significantly degrade communication-link performance. In \cite{Kodheli2021shunbian}, the authors introduced these coupled physical-layer impairments to the system analysis. To overcome these limitations and to fully exploit the limited onboard payload resources, advanced resource allocation paradigms are urgently required~\cite{wei2024fanshi}.

Up to now, non-orthogonal multiple access (NOMA) has been widely investigated for satellite Internet of Things (IoT) applications owing to its superior spectral efficiency and massive connectivity relative to orthogonal multiple access (OMA)~\cite{chun2021nomahao}. The concept of relay-aided LEO constellation NOMA systems was proposed in~\cite{zhang2025relay}, in which transmission reliability and coverage were enhanced to broaden the satellite service footprint. Furthermore, the authors of~\cite{yue2025federated} investigated the active simultaneously transmitting and reflecting surface assisted frameworks in terms of overcoming propagation blockages and enhancing connectivity in non-terrestrial networks (NTN). Focusing on resource allocation under spectrum scarcity,~\cite{hu2023noma} evaluated NOMA access schemes in satellite IoT networks, with particular attention to heterogeneous traffic accommodation. In multibeam satellite systems, the authors of~\cite{van2023joint} showed that joint power allocation and user scheduling can improve both fairness and throughput. In addition, the impact of channel ordering and successive interference cancellation (SIC) on NOMA in satellite scenarios was systematically analyzed in~\cite{dong2024xingnengfenxi}. Nevertheless, extending these static NOMA designs to VLEO environments raises severe challenges associated with dynamic topology and physical-layer impairments~\cite{bel2024noma}.

To address these challenges, coordinated transmission provides an effective way to improve spectral efficiency and service continuity in multi-satellite systems~\cite{shang2026duoxing,xuyang2026gnn,zhao2023duoxinge}. The central idea of multi-satellite coordination is to cooperatively serve users while flexibly allocating communication resources across both space and time~\cite{lin2024cordinate}. The employment of multi-satellite coordination has the following advantages over single-satellite transmission: i) Coordinated transmission introduces additional degrees of freedom that help maintain coverage continuity~\cite{pan2026mutist}; ii) The robustness of data-delivery can be extended to highly dynamic VLEO constellations~\cite{niu2025vleo}; and iii) Selection diversity can be designed effectively, which is particularly beneficial for mitigating abrupt service degradation during satellite handovers~\cite{wang2025duoxinge}. From the perspectives of fundamentals and system design, the authors of~\cite{shang2026duoxing} introduced multi-satellite cooperative communication for 6G, providing a careful comparison of practical applications and architectures. Building on these results, the authors in~\cite{xuyang2026gnn} analyzed the secrecy rate optimization in multi-satellite collaborative systems. To meet diverse requirements, the authors of~\cite{zhao2023duoxinge} studied multi-satellite cooperative communication by exploiting time asynchrony in non-orthogonal transmissions. In~\cite{yan2018pinpu}, cooperative NOMA in hybrid satellite-terrestrial relay networks was investigated, evaluating its advantages and weaknesses relative to conventional orthogonal designs. Despite these advances, practical implementation in highly dynamic VLEO environments remains challenging~\cite{crisp2020benefits}.

In response to the prohibitive computational complexity associated with coordinated alignment and joint resource scheduling, beam hopping (BH) has emerged as an effective transmission strategy for multibeam satellite systems~\cite{han2024bh}. To enable flexible spatio-temporal coverage, the authors of~\cite{chen2024bhxin1} proposed selectively activating only a subset of coordinated beams in each time slot, allowing satellite resources to be dynamically concentrated on high-demand regions. Such a mechanism is particularly well suited to accommodating heterogeneous traffic distributions~\cite{wang2026duoboshu}, where stringent onboard power constraints must be simultaneously satisfied. Moreover, the authors in~\cite{chen2023bhge} demonstrated that BH offers superior coverage flexibility compared with conventional full-illumination schemes. Throughput-oriented BH scheduling strategies have been extensively investigated in recent years~\cite{zhang2026uoboshu}, which aim to further improve spectral efficiency and overall system throughput. In addition, the authors of~\cite{jia2025bhxin2} highlighted that BH facilitates the matching of scarce resources to uneven spatial-temporal service demands in NTN. Nevertheless, the practical effectiveness of BH is closely tied to channel variation, switching accuracy, and user distribution~\cite{zhang2025research}, rendering its design substantially more challenging in highly dynamic communication environments. In~\cite{xuyang2023nomabh}, the authors explored the integration of NOMA assisted BH (NOMA-BH) in conventional LEO multibeam satellite systems. More recently, the limitations of extending such static or low-dynamic designs to VLEO networks were discussed in~\cite{luo2024vleo}. These limitations motivate a further step from NOMA-BH toward a NOMA assisted coordinated BH (NOMA-CBH) scheme with VLEO-specific propagation impairments.

\subsection{Motivation and Contributions}
As previously mentioned, existing studies on NOMA-BH mainly focus on conventional LEO systems and typically adopt relatively simplified channel assumptions. For instance, the authors in~\cite{xuyang2023nomabh} investigated the user throughput of NOMA-BH under static channel conditions to improve beam hopping efficiency. A recent study in~\cite{zhao2025changguilos} considered a multi-satellite LEO beam-hopping system under clear-sky propagation, where the Doppler frequency offset was assumed to be compensated and rain attenuation was neglected. Although these studies provide important foundations for NOMA-based beam scheduling and resource allocation, their direct application to VLEO systems requires further consideration of the rapid geometric variations and physical-layer impairments associated with very low orbital altitudes. In highly dynamic VLEO environments, CBH can effectively maintain continuous coverage by dynamically activating beams and allocating resources across spatio-temporal regions. To the best of our knowledge, the performance of NOMA-CBH under VLEO-specific channel conditions has not been investigated yet. Inspired by the limitations of existing studies, we investigate the NOMA-CBH scheme in dynamic VLEO environments by formulating a joint resource optimization problem to maximize the sum rate. To address this problem, we design a low-complexity decoupled (LCD) algorithm for beam scheduling and power allocation. Furthermore, under the resulting transmission structure, we analyze the outage performance by explicitly accounting for the considered VLEO channel characteristics. According to the aforementioned explanations, the main contributions of this work are summarized as follows:
\begin{itemize}
    \item[1)] We design a NOMA-CBH scheme for VLEO satellite communication systems and formulate a system sum rate maximization problem under the constraints of quality-of-service (QoS) requirements and the number of active beams. To address this non-convex problem, we design an efficient suboptimal LCD algorithm, which sequentially performs beam scheduling and power allocation through two cascaded stages, thereby reducing the computational complexity of the joint optimization.
    \item[2)] We establish a VLEO-oriented channel model that accounts for rapid geometry variations and relevant physical-layer impairments. In addition to geometry-dependent path variations, the model incorporates residual Doppler effects, atmospheric attenuation, and pointing perturbations, thereby providing a unified physical-layer basis for the design and evaluation of the considered NOMA-CBH scheme.
    \item[3)] We derive the closed-form expressions for the outage probabilities of the ground users and station. To gain further insights, we derive the corresponding asymptotic outage probabilities and obtain the diversity orders. Furthermore, the system throughput under delay-limited transmission is evaluated based on the derived outage probabilities. Simulation results confirm that the designed NOMA-CBH scheme achieves substantial improvements in system sum rate and outage performance compared to conventional orthogonal baselines.
\end{itemize}

\subsection{Organization and Notations}
The rest of this paper is organized as follows. Section II introduces the system model and the considered VLEO-specific channel model. Section III presents the joint problem formulation along with the designed LCD algorithm. In Section IV, we present the outage probability analysis and the corresponding asymptotic results. Section V provides numerical results to validate the effectiveness of the designed NOMA-CBH scheme, and the work is concluded in Section VI.

The main notations used throughout this paper are summarized as follows. $\mathbb{E}\{\cdot\}$ denotes the expectation operator. $\Gamma(\cdot)$ is the Gamma function, and $\binom{n}{k}$ denotes the binomial coefficient. $|\cdot|$ and $|\cdot|^2$ represent the modulus and squared modulus, respectively, while $(\cdot)^T$ denotes the transpose operator. $\Omega$ denotes the average channel power gain. Moreover, $\mathcal{A}$ denotes the discrete power-allocation search set of size $N_a$, and $\mathcal{A}^*$ denotes the set of selected slot-wise optimal power-allocation vectors.

\section{System Model}\label{System Model}
\begin{figure}[t!]
    \begin{center}
        \includegraphics[width=3.6in]{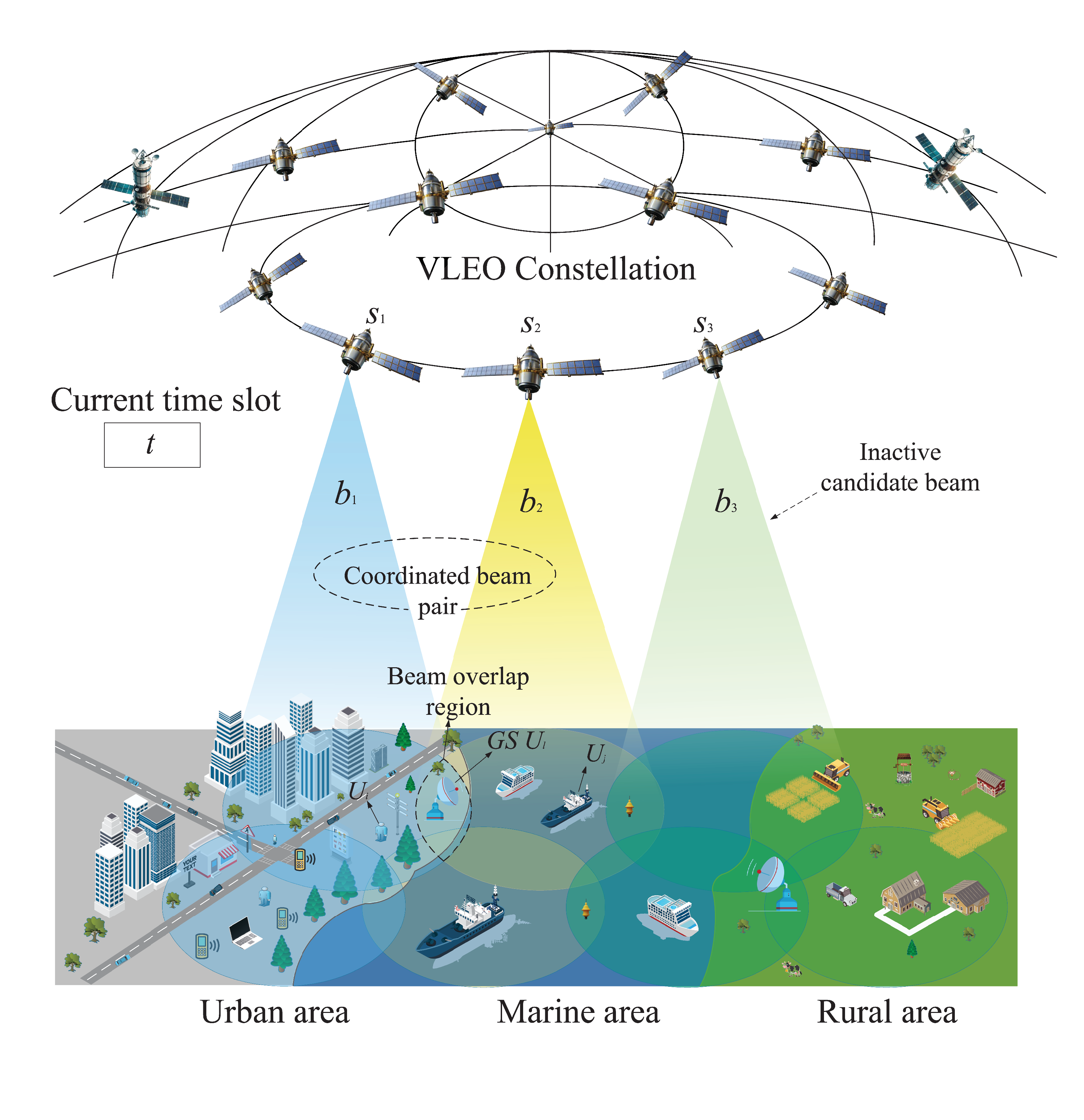}
        \caption{System model of the NOMA assisted VLEO satellite CBH communication network.}
        \label{Fig. 1.}
    \end{center}
\end{figure}
We consider NOMA assisted VLEO satellite CBH communication scenarios, as shown in Fig.~1, where multiple satellites coordinate their active beams to enable multidimensional transmission across geographically heterogeneous environments, including urban, marine, and rural regions~\cite{xuyang2026secure}. Let $\mathcal{S}=\{1,\dots,S\}$ denote the set of VLEO satellites, where each satellite $s \in \mathcal{S}$ is equipped with a set of candidate beams $\mathcal{B}_s$. The overall candidate-beam set is denoted by $\mathcal{B}=\bigcup_{s\in\mathcal{S}}\mathcal{B}_s$, where the satellite-specific beam sets are mutually disjoint, ensuring that each beam uniquely belongs to a specific satellite. A pool of $M$ ground users is distributed. The $M$ candidate ground users constitute the set $\mathcal{M}$, while the dedicated ground station (GS), denoted by $U_l$ and indexed by $l$, is treated separately from $\mathcal{M}$. In each transmission instance, the GS is mandatorily included in the NOMA-CBH service set, while two ordinary ground users are selected from the user pool. After the beam-user association is determined, the two selected users are ordered according to their instantaneous channel strengths and relabeled as $i$ and $j$ for SIC decoding. Time is discretized into slots indexed by $t$, and BH is employed for dynamic coverage and resource scheduling. In each slot, only a subset of beams is activated to illuminate selected geographical regions, while the remaining beams are kept silent. In the considered NOMA-CBH transmission, two beams from distinct satellites are jointly activated as a coordinated beam pair. The selected ground users are individually served by their associated satellite beams, whereas the GS is jointly served by the coordinated beam pair. Accordingly, beam activation and beam-user association constitute the main scheduling operations of the considered scheme.

To support slot-wise coordination, a logical network-side scheduler located at the ground gateway/network-control center determines the beam-user association and power-allocation decisions based on the available large-scale channel and geometry information. 
The participating satellites exchange the resulting scheduling and synchronization-control information through low-rate inter-satellite links (ISLs) before each transmission slot.
More specifically, the exchanged control information includes the selected satellite-beam/user associations, power-allocation coefficients, slot timing, and synchronization-control parameters.
The coordination information is updated once per transmission slot, i.e., every $T_{\mathrm{slot}}$, which is $0.1$~s in the numerical evaluation.
The corresponding payload blocks are delivered through the feeder/backhaul network and buffered at both satellites in advance so that they can transmit the same NOMA superposition.
The coordination delay is assumed to remain below the slot duration $T_{\mathrm{slot}}$; if the coordination delay exceeds this deadline or the required ISL connectivity is unavailable, the corresponding dual-satellite transmission opportunity is regarded as unavailable for that slot rather than reusing an outdated coordination decision.
Residual synchronization imperfections are captured by the residual Doppler loss and the MRC combining-loss factor introduced in Section~II-A.
Under the above latency assumption, the scheduler uses the latest available large-scale geometry and channel information at the beginning of each slot, while the additional performance degradation caused by explicitly delayed or stale geometry/CSI is not separately modeled in the present work.

\subsection{VLEO Channel Model}
The downlink transmission is primarily characterized by large-scale fading. Let $h_{s,b,m,t}$ denote the channel from beam $b \in \mathcal{B}_s$ of satellite $s$ to user $m$ in time slot $t$. This coefficient incorporates the distinctive features of the VLEO environment, including geometric dynamics, attitude perturbations, and free-space path loss. Consequently, the channel model is explicitly formulated as
\begin{equation}
\label{eq:channel_gain}
\begin{split}
h_{s,b,m,t} = {} & \eta_{s,b,m,t} \sqrt{G_r} \left( \frac{\lambda}{4\pi d_{s,m,t}} \right) \\
& \times \sqrt{ G_t\!\left(\phi^{\rm eff}_{s,b,m,t}\right) L_{{\rm Dop},s,m,t} L_{\rm atm} } e^{j\Psi_{s,m,t}},
\end{split}
\end{equation}
where
\begin{itemize}
    \item $\eta_{s,b,m,t}$ denotes the small-scale fading coefficient, whose envelope follows a Nakagami-$\mu$ distribution, $G_r$ represents the receive antenna gain, and $\lambda$ is the carrier wavelength. Here $G_r$ takes the user receive gain for the ground users and is replaced by the GS receive gain $G_l$ for the GS. Distinct from traditional satellite channel models, the designed model explicitly accounts for dynamic impairments specific to VLEO.
    \item $d_{s,m,t}$: The instantaneous slant range is given by $d_{s,m,t} = \sqrt{r_{s,t}^2 + R_E^2 - 2 r_{s,t} R_E \cos(\vartheta_{s,m,t})}$, which varies rapidly due to the high orbital velocity~\cite{zhang2025xindao}. In this expression, $r_{s,t}$, $R_E$, and $\vartheta_{s,m,t}$ represent the geocentric distance of satellite $s$ in time slot $t$, Earth radius, and the corresponding central angle between satellite $s$ and user $m$, respectively.
    \item $\phi^{\rm eff}_{s,b,m,t}$ denotes the effective off-axis angle of beam $b$ from satellite $s$ to terminal $m$ at slot $t$. Let $\phi^{\rm geo}_{s,b,m,t}$ denote the corresponding nominal geometry-dependent off-axis angle. Considering two-dimensional pointing perturbations, we have $\phi^{\rm eff}_{s,b,m,t} = \sqrt{\left(\phi^{\rm geo}_{s,b,m,t} + \delta_{\alpha,s,t}\right)^2 + \delta_{\varpi,s,t}^{2}}$, where $\delta_{\alpha,s,t}$ and $\delta_{\varpi,s,t}$ are independent zero-mean Gaussian pointing errors with standard deviation $\sigma_p$. The resulting pointing impairment is captured by $G_t(\phi^{\rm eff}_{s,b,m,t})$.
    \item $L_{\mathrm{atm}}$: The term $L_{\mathrm{atm}} = 10^{-L_{\alpha}(\theta_{s,m,t})/10}$ accounts for geometry-dependent atmospheric absorption, with $L_{\alpha}(\theta_{s,m,t}) = \frac{L_{\mathrm{zen}}}{\sin \theta_{s,m,t}}$ representing the path-dependent loss derived from the zenith attenuation $L_{\mathrm{zen}}$ and the user elevation angle $\theta_{s,m,t}$. The instantaneous Doppler shift is determined by the satellite--terminal slant-range rate. Under the adopted circular-orbit geometry, the instantaneous range rate is given by $\dot d_{s,m,t} = \frac{R_E v_{\mathrm{sat}} \sin(\vartheta_{s,m,t})}{d_{s,m,t}}$, and the corresponding Doppler shift is $f_{d,s,m,t} = \frac{\dot d_{s,m,t}}{\lambda} = \frac{f_c}{c}\dot d_{s,m,t}$, where $\lambda=c/f_c$ is the carrier wavelength. The sign of $f_{d,s,m,t}$ indicates whether the satellite is approaching or receding from the terminal. After Doppler compensation, the residual frequency offset is modeled as a fixed fraction of the instantaneous Doppler magnitude, i.e., $\Delta f_{\mathrm{res},s,m,t} = \epsilon_{\mathrm{Dop}} \left|f_{d,s,m,t}\right|$, where $\epsilon_{\mathrm{Dop}}=0.05$ denotes the residual Doppler factor.
    \item $\Psi_{s,m,t}$: To capture the high-mobility characteristic of VLEO systems, $\Psi_{s,m,t} = \int_{0}^{t T_{\mathrm{slot}}} 2\pi f_{d,s,m}(\tau) d\tau$ denotes the cumulative Doppler phase rotation induced by the instantaneous Doppler shift $f_{d,s,m}(\tau)$, where $T_{\mathrm{slot}}$ denotes the slot duration.
    \item $L_{\mathrm{Dop},s,m,t}$: In practical high-mobility transmissions, Doppler compensation is typically performed at the receiver. Nevertheless, imperfect carrier synchronization may leave a residual frequency offset, leading to coherent signal-power loss over the effective observation interval. To characterize this post-compensation degradation, we introduce a residual Doppler loss factor $L_{\mathrm{Dop},s,m,t} = \mathrm{sinc}^2(\Delta f_{\mathrm{res},s,m,t} T_{\mathrm{eff}})$, where $\Delta f_{\mathrm{res},s,m,t}$ and $T_{\mathrm{eff}}$ denote the residual frequency offset and the effective coherence time, respectively~\cite{zuo2023canyudoppler}. For the dual-satellite transmission toward the GS $U_l$, the two received branches are assumed to be time-aligned and carrier-frequency synchronized within the receiver synchronization tolerance. The remaining link-dependent frequency offsets are captured by $\Delta f_{\mathrm{res},s,l,t}$, while the branch phases are estimated and compensated by the MRC receiver.
\end{itemize}

Additionally, $L_{{\rm Dop},s,m,t}$ and $L_{\rm atm}$ are dimensionless loss factors incorporated into the large-scale channel gain, while the pointing-induced attenuation is captured through the effective off-axis angle $\phi^{\rm eff}_{s,b,m,t}$ in the transmit antenna gain.

The VLEO channel model in \eqref{eq:channel_gain} incorporates slot-dependent geometric variations and Doppler-induced phase accumulation. Despite this, we adopt a large-scale block-fading model. The satellite position and the associated large-scale channel parameters are updated at the beginning of each time slot of duration $T_{\mathrm{slot}}$. Since the satellite displacement within one slot is small relative to the corresponding slant range, the large-scale fading is treated as quasi-static within slot $t$, while the rapid small-scale dynamics are absorbed statistically.

To quantitatively justify this approximation, the maximum satellite displacement within one slot is $v_{\mathrm{sat}}T_{\mathrm{slot}}=0.77$ km for $v_{\mathrm{sat}}=7.70$ km/s and $T_{\mathrm{slot}}=0.1$ s. Since the minimum slant range is approximately $360$ km, the corresponding relative range variation is conservatively bounded by $\frac{\Delta d}{d} \leq \frac{0.77}{360} \simeq 2.14\times10^{-3}$. Under the adopted circular-orbit geometry, the one-slot variations of the elevation angle and the geometry-dependent off-axis angle are both bounded by approximately $0.123^\circ$. The latter is only about $3.7\%$ of the $3$-dB beamwidth, $\theta_{3\mathrm{dB}}\simeq3.29^\circ$, obtained from the adopted aperture-antenna pattern. Moreover, over the considered visible orbital region, the one-slot variation of the instantaneous Doppler shift is below approximately $0.61$ kHz. With the residual Doppler factor $\epsilon_{\mathrm{Dop}}=0.05$, the corresponding residual-frequency variation is below approximately $30.5$ Hz, yielding $|\Delta f_{\mathrm{res}}|T_{\mathrm{eff}} < 3.7\times10^{-4} \ll 1$ for $T_{\mathrm{eff}}=12~\mu$s.

Accordingly, the quasi-static large-scale approximation is regarded as valid when
\begin{subequations}\label{eq:quasi_static_cond}
\begin{align}
\frac{\Delta d}{d} &\ll 1, \label{eq:cond_a} \\
\frac{\Delta\theta_{\mathrm{elev}}}{\theta_{\min}} &\ll 1, \label{eq:cond_b} \\
\frac{\Delta\phi^{\mathrm{geo}}}{\theta_{3\mathrm{dB}}} &\ll 1, \label{eq:cond_c} \\
|\Delta f_{\mathrm{res}}|T_{\mathrm{eff}} &\ll 1. \label{eq:cond_d}
\end{align}
\end{subequations}

The above bounds satisfy these conditions for the adopted $T_{\mathrm{slot}}=0.1$ s, thereby supporting the treatment of the large-scale channel parameters as constant within each slot.

As the theoretical analysis in Section IV depends only on the signal envelope, the phase rotation term $e^{j\Psi_{s,m,t}}$ does not affect the statistical distribution of the channel gain. In contrast, the numerical simulations in Section V incorporate the full dynamic characteristics, including Doppler effects, attitude perturbations, and atmospheric attenuation, to ensure rigorous validation. Furthermore, the satellite transmit antenna gain $G_t(\varphi)$ is expressed as
\begin{equation}
    \scalebox{0.91}{$\displaystyle
    G_t(\varphi) = \left\{
        \begin{array}{ll}
            G_{\max}, & \varphi = 0^\circ; \\[10pt]
            G_{\max} \left| \dfrac{2 J_1(2\pi \Xi \sin \varphi / \lambda)}{2\pi \Xi \sin \varphi / \lambda} \right|^2, & 0^\circ < |\varphi| \leq 90^\circ
        \end{array},
    \right.
    $}
\end{equation}
where $G_{\max}$ denotes the maximum main lobe gain of the satellite antenna, given by $G_{\max} = \left( \frac{2\pi\Xi}{\lambda} \right)^2$, and $\Xi$ represents the aperture radius of the satellite transmitting antenna. Additionally, $\varphi$ denotes the generic off-axis angle argument of the antenna pattern. For the link $(s,b,m,t)$ in \eqref{eq:channel_gain}, the antenna gain is evaluated at $\varphi=\phi^{\rm eff}_{s,b,m,t}$. Moreover, $J_1(\cdot)$ denotes the first-order Bessel function of the first kind.

The beam footprint is represented implicitly through the geometry-dependent aperture-antenna gain $G_t\!\left(\phi_{s,b,m,t}^{\mathrm{eff}}\right)$, rather than by an additional hard-edged coverage boundary. A terminal is regarded as geometrically serviceable only when the minimum-elevation requirement is satisfied, while its relative location within the beam footprint is captured by the corresponding off-axis antenna gain.

\subsection{Signal Model}
In the considered system configuration, $N_B=2$ beams are simultaneously activated in each slot, and these two active beams constitute the coordinated NOMA-CBH beam pair. Hence, no additional non-coordinated cochannel beam is simultaneously active in the evaluated transmission instance. The two coordinated beams fully reuse the same frequency resource and transmit the same NOMA superposition, whose intra-cluster interference is explicitly captured by the SIC-based SINR expressions below. Accordingly, additional inter-beam interference from non-coordinated active beams is not included in the present model. Therefore, the received signals at these terminals can be formulated as
\begin{align}
y_{s,b,w,t} = h_{s,b,w,t}\left(\sqrt{P_i}x_i+\sqrt{P_j}x_j+\sqrt{P_l}x_l\right)+n_w,
\label{jieshouxinhao}
\end{align}
where $n_w \sim \mathcal{CN}(0,\sigma^2)$ denotes the additive white Gaussian noise at terminal $w$, and $x_i$, $x_j$, and $x_l$ denote the signals intended for users $U_i$, $U_j$, and $U_l$, respectively. For each ground user, $(s,b)$ denotes its associated serving satellite-beam pair. For the GS $U_l$, \eqref{jieshouxinhao} applies separately to the two coordinated satellite-beam branches. The two coordinated beams transmit the same NOMA superposition. For each ordinary user, the non-associated coordinated-beam component is assumed to be sufficiently attenuated by the off-axis transmit-beam pattern and is therefore neglected, while the two spatially selective receive chains at the GS resolve the two serving-satellite branches before MRC. Let $P_{\mathrm{tot}}$ denote the total transmit power allocated to each active beam, and let $a_{i,t}$, $a_{j,t}$, and $a_{l,t}$ denote the corresponding power allocation coefficients, satisfying $a_{i,t}+a_{j,t}+a_{l,t}=1$. The two coordinated beams employ the same NOMA power-allocation coefficients. Then, the power allocated to signal $x_w$ is given by $P_w=a_{w,t} P_{\mathrm{tot}}$. Accordingly, the signal-to-interference-plus-noise ratio (SINR) at $U_i$ is formulated as~\cite{yue2025yinbitongxin}
\begin{align}
\gamma_i = \frac{|h_{s,b,i,t}|^2 P_i}{|h_{s,b,i,t}|^2 P_j + |h_{s,b,i,t}|^2 P_l + \sigma^2}.
\label{izhongduan}
\end{align}

Similarly, the SINR for $U_j$ when decoding the signal intended for $U_i$ can be expressed as
\begin{align}
\gamma_{j \to i} = \frac{|h_{s,b,j,t}|^2 P_i}{|h_{s,b,j,t}|^2 P_j + |h_{s,b,j,t}|^2 P_l + \sigma^2}.
\label{jjiancei}
\end{align}

With the interference from $U_i$ successfully canceled, the SINR for $U_j$ to decode its own message $x_j$ is given by
\begin{align}
\gamma_j = \frac{|h_{s,b,j,t}|^2 P_j}{|h_{s,b,j,t}|^2 P_l + \sigma^2}.
\label{jjiancezishen}
\end{align}

Subsequently, for $U_l$, the SINR after successively canceling the interference from $U_i$ and $U_j$ through the SIC process is formulated as
\begin{align}
\gamma_{l \to i} = \frac{|h_{l,t}^{\mathrm{MRC}}|^2 P_i}{|h_{l,t}^{\mathrm{MRC}}|^2 P_j + |h_{l,t}^{\mathrm{MRC}}|^2 P_l + \sigma^2},
\label{ljiancei}
\end{align}
and
\begin{align}
\gamma_{l \to j} = \frac{|h_{l,t}^{\mathrm{MRC}}|^2 P_j}{|h_{l,t}^{\mathrm{MRC}}|^2 P_l + \sigma^2}.
\label{ljiancej}
\end{align}

Finally, the SINR for $U_l$ to decode its own signal is expressed as
\begin{align}
\gamma_l = \frac{|h_{l,t}^{\mathrm{MRC}}|^2 P_l}{\sigma^2},
\label{ljiancezishen}
\end{align}
where $|h_{l,t}^{\mathrm{MRC}}|^2 = \chi (|h_{s_1,b_1,l,t}|^2 + |h_{s_2,b_2,l,t}|^2)$ represents the post-combining channel gain at the GS, where $b_1 \in \mathcal{B}_{s_1}$ and $b_2 \in \mathcal{B}_{s_2}$ denote the two coordinated beams belonging to two distinct satellites. The GS is equipped with two receive chains and performs MRC over the two coordinated beams. The factor $\chi \in (0,1]$ captures the post-combining loss induced by imperfect channel estimation and residual carrier synchronization under the rapid Doppler variations of VLEO links. Under the symmetric large-scale fading assumption for the two independent Nakagami-$\mu$ links, the post-combining channel power follows a Gamma distribution, expressed as $|h_{l,t}^{\mathrm{MRC}}|^2 \sim \Gamma(2\mu,\zeta_l)$, where the scale parameter $\zeta_l$ includes the MRC combining loss. When the large-scale gains of the two coordinated links differ, the composite power can be approximated by a single equivalent Gamma distribution via second-order moment matching for analytical tractability~\cite{jung2024mrc}. To account for all required SIC decoders, the effective decoding SINRs of the weak and intermediate users are defined as $\gamma_i^{\mathrm{eff}} = \min\{\gamma_i, \gamma_{j\rightarrow i}, \gamma_{l\rightarrow i}\}$ and $\gamma_j^{\mathrm{eff}} = \min\{\gamma_j, \gamma_{l\rightarrow j}\}$, respectively, while $\gamma_l^{\mathrm{eff}} = \gamma_l$ for the station. Considering both beam activation and beam-user association, the instantaneous achievable spectral efficiency and transmission rate of user $m$ in time slot $t$ are formulated as
\begin{align}
R_{m,t}^{\mathrm{SE}} = \sum_{s \in \mathcal{S}, b \in \mathcal{B}_s} z_{s,b,t} \beta_{s,b,m,t} \log_2(1 + \gamma_m^{\mathrm{eff}}),
\label{eq:achievable_se}
\end{align}
and
\begin{align}
R_{m,t} = B_{\max} R_{m,t}^{\mathrm{SE}},
\label{eq:achievable_rate}
\end{align}
where $R_{m,t}^{\mathrm{SE}}$ represents the corresponding spectral efficiency in BPCU, and $B_{\max}$ denotes the system bandwidth. The binary variables $z_{s,b,t}$ and $\beta_{s,b,m,t}$ denote the beam-activation and beam-user-association indicators, respectively, and are formally defined in Section~III-A. For the mandatory GS $U_l$, its instantaneous spectral efficiency and transmission rate are respectively given by $R_{l,t}^{\mathrm{SE}}=\log_2(1+\gamma_l^{\mathrm{eff}})$ and $R_{l,t}=B_{\max}R_{l,t}^{\mathrm{SE}}$. Furthermore, satellite orbital data and the corresponding Doppler frequency shifts are incorporated into the mathematical model used in the subsequent algorithm.

\section{PROBLEM FORMULATION AND ALGORITHM DESIGN}\label{Section_III}
\subsection{Problem Formulation}
In order to enhance spectral efficiency and resource utilization, we formulate a joint resource optimization problem that focuses on the joint design of beam scheduling and power allocation, with the objective of maximizing the system sum rate. The beam activation variable is defined as
\begin{equation}
\scalebox{0.86}{$
z_{s,b,t} = \begin{cases}
1, & \text{beam } b \text{ of satellite } s \text{ is activated in time slot } t; \\
0, & \text{otherwise}
\end{cases}
$}.
\end{equation}

To explicitly characterize the beam-user association, we introduce a binary matching variable $\beta_{s,b,m,t}$, where $\beta_{s,b,m,t}=1$ indicates that user $m$ is served by beam $b$ of satellite $s$ in time slot $t$, and $\beta_{s,b,m,t}=0$ otherwise. The system sum rate is maximized by jointly optimizing the beam scheduling variables, i.e., the beam-activation variables $z_{s,b,t}$ and the beam-user association variables $\beta_{s,b,m,t}$, together with the power-allocation coefficients $a_{w,t}$ for the scheduled NOMA group. Accordingly, the joint optimization problem is formulated as

\begingroup
\fontsize{8.8pt}{10.0pt}\selectfont
\begin{subequations} \label{eq:problem_P0}
\begin{align}
\mathcal{P}_0: \quad & \max_{z_{s,b,t}, \beta_{s,b,m,t}, a_{w,t}} \sum_{t \in \mathcal{T}} \left( \sum_{m \in \mathcal{M}} R_{m,t} + R_{l,t} \right) \label{eq:objective_func} \\
\text{s.t.} \quad & \sum_{w\in\{i,j,l\}} a_{w,t} = 1, \forall t \in \mathcal{T}, \label{eq:const_power_sum} \\
& a_{l,t} \leq a_{j,t} \leq a_{i,t}, \forall t \in \mathcal{T}, \label{eq:const2} \\
& \sum_{s \in \mathcal{S}, b \in \mathcal{B}_s} z_{s,b,t} = N_B, \forall t \in \mathcal{T}, \label{eq:const3} \\
& \beta_{s,b,m,t} \leq z_{s,b,t}, \forall s \in \mathcal{S}, \forall b \in \mathcal{B}_s, \forall m \in \mathcal{M}, \forall t \in \mathcal{T}, \label{eq:const_x1} \\
& \sum_{s \in \mathcal{S}, b \in \mathcal{B}_s} \beta_{s,b,m,t} \leq 1, \forall m \in \mathcal{M}, t \in \mathcal{T}, \label{eq:const_x2} \\
& 0 \leq a_{w,t} \leq 1, \forall w \in \{i,j,l\}, \forall t \in \mathcal{T}, \label{eq:const_aw_range} \\
& R_{m,t}^{\mathrm{SE}} \geq R_m^{\min} \sum_{s \in \mathcal{S}, b \in \mathcal{B}_s} \beta_{s,b,m,t}, \forall m \in \mathcal{M}, \forall t \in \mathcal{T}, \label{eq:const5} \\
& R_{l,t}^{\mathrm{SE}} \geq R_l^{\min}, \forall t \in \mathcal{T}, \label{eq:const_rl_min} \\
& \sum_{m \in \mathcal{M}} \beta_{s,b,m,t} = z_{s,b,t}, \forall s \in \mathcal{S}, \forall b \in \mathcal{B}_s, \forall t \in \mathcal{T}, \label{eq:const_beta_z} \\
& \sum_{b \in \mathcal{B}_s} z_{s,b,t} \le 1, \forall s \in \mathcal{S}, \forall t \in \mathcal{T}, \label{eq:const_single_beam} \\
& \gamma_{j\rightarrow i} \geq \gamma_i^{\mathrm{th}}, \gamma_{l\rightarrow i} \geq \gamma_i^{\mathrm{th}}, \gamma_{l\rightarrow j} \geq \gamma_j^{\mathrm{th}}, \label{eq:const_sic}
\end{align}
\end{subequations}
\endgroup
where \eqref{eq:const_power_sum} ensures that the normalized power-allocation coefficients of the scheduled NOMA group sum to unity. Here, $\gamma_w^{\mathrm{th}} = 2^{R_w^{\min}} - 1$ denotes the target SINR corresponding to the minimum spectral-efficiency requirement of terminal $U_w$. Constraint \eqref{eq:const2} specifies the power-ordering relation among the multiplexed terminals, while the explicit decoding constraints in \eqref{eq:const_sic} are imposed for each scheduled group to ensure SIC feasibility. For beam scheduling, \eqref{eq:const3} ensures that $N_B$ coordinated beams are activated in each time slot. Together with \eqref{eq:const_single_beam}, this guarantees that the two coordinated active beams originate from two distinct satellites. The constraints for $\beta_{s,b,m,t}$ are governed by \eqref{eq:const_x1} and \eqref{eq:const_x2}, which ensure that users are associated only with active beams and that each user is served by at most one beam per time slot. Constraint \eqref{eq:const_aw_range} gives the feasible range of each power-allocation coefficient. To satisfy the QoS requirements, \eqref{eq:const5} requires $R_{m,t}^{\mathrm{SE}}$ of each scheduled user to be no less than the minimum spectral-efficiency threshold $R_m^{\min}$. Constraint \eqref{eq:const_rl_min} guarantees the minimum spectral-efficiency requirement of the mandatory GS $U_l$. Constraint \eqref{eq:const_beta_z} ensures that each active beam is associated with exactly one ordinary ground user.

Further analysis reveals a strong coupling among $z_{s,b,t}$, $\beta_{s,b,m,t}$, and $a_{w,t}$. More precisely, once the coordinated beams have been activated, appropriate terminal matching must first be determined according to the large-scale channel conditions, after which $a_{w,t}$ must be carefully adjusted to guarantee successful SIC and satisfy the QoS constraints of the users. As a result, jointly optimizing the binary scheduling decisions, the matching relations, and the continuous power variables yields a mixed-integer nonlinear programming (MINLP) problem of prohibitive complexity. Motivated by this structural property, we design the LCD algorithm that decomposes $\mathcal{P}_0$ into two cascaded subproblems:
\begin{enumerate}
    \item A beam scheduling and beam-user association subproblem driven by large-scale channel gains.
    \item A NOMA power allocation subproblem for the scheduled group.
\end{enumerate}

By addressing these two stages sequentially, the computational burden is substantially reduced at the cost of a possible slot-wise feasibility limitation and a quantifiable suboptimality gap, the latter of which is evaluated through the exhaustive discrete benchmark in Section~V-A.

\subsection{Solving the Beam Scheduling and Beam-User Association Subproblem}
We first address the non-convexity of $\mathcal{P}_0$. Since $\mathcal{P}_0$ involves coupled binary and continuous variables, it constitutes an MINLP problem, which is generally difficult to solve optimally~\cite{deng2008lao}. In VLEO scenarios, short service-visibility windows and practical onboard processing constraints motivate low-complexity resource scheduling. Accordingly, we adopt a decoupling strategy that yields an efficient suboptimal solution suitable for real-time scheduling.

Based on the preceding analysis, the subproblem for $z_{s,b,t}$ and $\beta_{s,b,m,t}$ is formulated to maximize the large-scale scheduling metric $G^{\rm LS}_{s,b,m,t}$, subject to system constraints. This large-scale gain is given by $G^{\rm LS}_{s,b,m,t} = G_r G_t\!\left(\phi^{\rm eff}_{s,b,m,t}\right) \left(\frac{\lambda}{4\pi d_{s,m,t}}\right)^2 L_{{\rm Dop},s,m,t} L_{\rm atm}$, which incorporates the instantaneous propagation distance, atmospheric attenuation, residual Doppler loss, and pointing perturbations through the effective off-axis angle.

For each time slot $t$, let $\mathcal{F}_t$ denote the set of admissible beam-user pairs $(s,b,m)$ satisfying the terminal-visibility and minimum-elevation requirements. Moreover, let $\mathcal{C}_t$ denote the set of structurally admissible coordinated beam pairs formed by beams from distinct satellites. Accordingly, the beam scheduling and beam-user association subproblem is formulated as
\begin{subequations} \label{eq:problem_P1}
\begin{align}
\mathcal{P}_1: \quad
&\max_{z_{s,b,t}, \beta_{s,b,m,t}}
\sum_{s \in \mathcal{S}, b \in \mathcal{B}_s,m \in \mathcal{M},t \in \mathcal{T}}
\beta_{s,b,m,t} G_{s,b,m,t}^{LS}
\\
&\text{s.t.} \quad
\eqref{eq:const3}, \eqref{eq:const_x1}, \eqref{eq:const_x2},
\eqref{eq:const_beta_z}, \eqref{eq:const_single_beam},
\label{eq:P1_structural_constraints}
\\
& \beta_{s,b,m,t}=0,
\forall (s,b,m)\notin\mathcal{F}_t,\ \forall t\in\mathcal{T},
\label{eq:P1_feasible_pair}
\\
& \left\{(s,b):z_{s,b,t}=1\right\}\in\mathcal{C}_t,
\forall t\in\mathcal{T}.
\label{eq:P1_coordinated_pair}
\end{align}
\end{subequations}

This criterion, based on large-scale gain, transforms the temporal beam scheduling and beam-user association problem into a large-scale-gain-driven scheduling problem within each time slot. Due to the short small-scale coherence time in VLEO systems, circumventing instantaneous small-scale channel state information (CSI) in the scheduling stage and driving scheduling by large-scale gains contracts the search space and reduces the real-time computational burden. Consequently, based on the slot-dependent large-scale gain matrix $\mathbf{G}_t = [G_{s,b,m,t}^{\mathrm{LS}}]_{(s,b),m}$, the binary beam-user association decision $\mathbf{Z}_t^*$ is determined as
\begin{equation}
\mathbf{Z}_t^* \triangleq \left[\beta_{s,b,m,t}^*\right]_{(s,b),m},
\end{equation}
and the corresponding beam-activation decision is given by
\begin{equation}
z_{s,b,t}^* = \sum_{m \in \mathcal{M}} \beta_{s,b,m,t}^*.
\end{equation}

The admissibility constraint in \eqref{eq:P1_feasible_pair} excludes beam-user pairs that violate the terminal-visibility or minimum-elevation requirements, while \eqref{eq:P1_coordinated_pair} restricts the two active beams to a structurally admissible coordinated pair. Together with $N_B=2$ and \eqref{eq:const_beta_z}, exactly two ordinary ground users are scheduled in each time slot, while the dedicated GS $U_l$ is mandatorily included in the resulting NOMA-CBH group. The resulting association decisions are recorded in the sparse binary scheduling matrix $\mathbf{Z}_t^*$.

\begin{algorithm}
\caption{Time-Domain Beam Scheduling and NOMA Beam-User Association}
\begin{algorithmic}[1]
\renewcommand{\algorithmicrequire}{\textbf{Input:}}
\renewcommand{\algorithmicensure}{\textbf{Output:}}
\Require Satellite set $\mathcal{S}$, satellite-specific candidate beam sets $\{\mathcal{B}_s\}_{s\in\mathcal{S}}$, set of $M$ ground users $\mathcal{M}$, dedicated GS $U_l$, number of time slots $T$, number of coordinated active beams $N_B$, MRC combining-loss factor $\chi$, and orbital parameters.
\For{each time slot $t = 1$ to $T$}
    \State Update the positions of satellites $s\in\mathcal{S}$ and the corresponding beam and ground-terminal geometry.
    \For{each $s\in\mathcal{S}$, $b\in\mathcal{B}_s$, and $m\in\mathcal{M}$}
        \State Calculate the instantaneous distance $d_{s,m,t}$, effective off-axis angle $\phi^{\rm eff}_{s,b,m,t}$, atmospheric attenuation $L_{\rm atm}$, and residual Doppler loss $L_{{\rm Dop},s,m,t}$.
        \State Compute the large-scale channel gain $G_{s,b,m,t}^{LS}$.
    \EndFor
    \State Construct the slot-dependent large-scale gain matrix $\mathbf{G}_t = [G_{s,b,m,t}^{LS}]_{(s,b),m} \in \mathbb{R}_{+}^{N_{\mathrm{tot}}\times M}$.
    \State Obtain a greedy solution to $\mathcal{P}_1$ by selecting beam-user pairs from $\mathcal{F}_t$ in descending order of $G_{s,b,m,t}^{LS}$, subject to constraints \eqref{eq:const3}, \eqref{eq:const_x1}, \eqref{eq:const_x2}, \eqref{eq:const_beta_z}, \eqref{eq:const_single_beam}, and the coordinated-beam admissibility constraint \eqref{eq:P1_coordinated_pair}, until $N_B$ pairs are selected, thereby obtaining $\beta_{s,b,m,t}^{*}$.
    \State Obtain the corresponding beam-activation decisions as $z_{s,b,t}^{*}=\sum_{m\in\mathcal{M}}\beta_{s,b,m,t}^{*}$.
    \State Let the two selected beam-user pairs be $(s_1,b_1,u_1)$ and $(s_2,b_2,u_2)$, where $s_1\neq s_2$, and mandatorily include the GS $U_l$.
    \State Calculate the instantaneous channel gains of the two selected ground users: $|h_{s_k,b_k,u_k,t}|^2 = |\eta_{s_k,b_k,u_k,t}|^2 G_{s_k,b_k,u_k,t}^{LS}$, $k\in\{1,2\}$.
    \State Calculate the two coordinated channel branches toward the GS according to \eqref{eq:channel_gain} and obtain the MRC-combined gain: $|h_{l,t}^{\mathrm{MRC}}|^2 = \chi\left(|h_{s_1,b_1,l,t}|^2 + |h_{s_2,b_2,l,t}|^2\right)$.
    \State Relabel the two selected ordinary users as $i$ and $j$ according to their equivalent channel strengths and form the scheduled NOMA group $\{i,j,l\}$.
    \State Store the corresponding channel states in $\mathcal{H}_t$.
\EndFor
\Statex Obtain the scheduling and association decisions together with the channel-state sets.
\Ensure $z_{s,b,t}^{*}$, $\beta_{s,b,m,t}^{*}$, $\mathcal{H}=\{\mathcal{H}_t\}_{t\in\mathcal{T}}$.
\end{algorithmic}
\end{algorithm}

\subsection{Solving the Power Allocation Subproblem}
\begin{algorithm}
\caption{Discrete Power Allocation Optimization for LCD Algorithm}
\begin{algorithmic}[1]
\renewcommand{\algorithmicrequire}{\textbf{Input:}}
\renewcommand{\algorithmicensure}{\textbf{Output:}}

\Require Scheduling and association decisions
$\{\mathbf{Z}_t^*\}_{t\in\mathcal{T}}$,
channel-state set $\mathcal{H}=\{\mathcal{H}_t\}_{t\in\mathcal{T}}$,
discrete power-allocation search set
$\mathcal{A}=\{\mathbf{a}^{(1)},\ldots,\mathbf{a}^{(N_a)}\}$,
total transmit power $P_{\mathrm{tot}}$,
and QoS thresholds
$\{R_i^{\min},R_j^{\min},R_l^{\min}\}$.

\State Initialize system total rate $R_{\mathrm{tot}}\leftarrow 0$.
\State Initialize the set of optimal power-allocation vectors
$\mathcal{A}^*\leftarrow\emptyset$.

\For{each time slot $t=1$ to $T$}

    \State Identify the scheduled NOMA group $\{i,j,l\}$
    and the corresponding channel states from
    $\mathbf{Z}_t^*$ and $\mathcal{H}_t$.

    \State Initialize the slot-wise optimal rate
    $R_t^*\leftarrow 0$.

    \State Initialize the slot-wise optimal power-allocation vector
    $\mathbf{a}_t^*\leftarrow\emptyset$.

    \For{$k=1,\ldots,N_a$}

        \State Set
        $\mathbf{a}^{(k)}
        =[a_i^{(k)},a_j^{(k)},a_l^{(k)}]$
        and allocate
        $P_w=a_w^{(k)}P_{\mathrm{tot}}$,
        $w\in\{i,j,l\}$.

        \State Calculate the SINRs and rates
        $(R_{i,t},R_{j,t},R_{l,t})$
        according to \eqref{izhongduan}--\eqref{eq:achievable_rate}.

        \State Calculate the current sum rate
        $R_{\mathrm{curr}}
        \leftarrow R_{i,t}+R_{j,t}+R_{l,t}$.

        \If{$\mathbf{a}^{(k)}$ satisfies
        \eqref{eq:const_power_sum}, \eqref{eq:const2}, \eqref{eq:const_aw_range}, \eqref{eq:const5}, \eqref{eq:const_rl_min}, and \eqref{eq:const_sic},
        and $R_{\mathrm{curr}}>R_t^*$}

            \State $R_t^*\leftarrow R_{\mathrm{curr}}$.
            \State $\mathbf{a}_t^*\leftarrow\mathbf{a}^{(k)}$.

        \EndIf

    \EndFor

    \If{$\mathbf{a}_t^*\neq\emptyset$}

        \State $R_{\mathrm{tot}}
        \leftarrow R_{\mathrm{tot}}+R_t^*$.

        \State Add $\mathbf{a}_t^*$ to $\mathcal{A}^*$.

    \Else

        \State Record slot $t$ as infeasible under the QoS and SIC constraints.

    \EndIf

\EndFor

\Statex Obtain the slot-wise optimal power-allocation vectors
and the system total rate.

\Ensure $\mathcal{A}^*$, $R_{\mathrm{tot}}$.

\end{algorithmic}
\end{algorithm}

In this subsection, we investigate the power-allocation stage conditioned on the predetermined beam scheduling and beam-user association decisions. Since the binary scheduling and association decisions $z_{s,b,t}^{*}$ and $\beta_{s,b,m,t}^{*}$ returned by Algorithm~1 are fixed, they are no longer optimization variables in the second stage. Conditioned on these scheduling decisions, the remaining power-allocation stage is formulated as a lower-dimensional continuous subproblem, denoted as $\mathcal{P}_2$. This subproblem maximizes the system sum rate over $a_{w,t}$ for the selected NOMA group, subject to the power-ordering, QoS, and SIC-feasibility constraints. The existence of a feasible power-allocation vector is not guaranteed a priori; if no candidate satisfies these constraints, the corresponding slot is recorded as infeasible according to Algorithm~2. The instantaneous sum rate in time slot $t$ is given by
\begin{equation}
R_{\mathrm{sum},t} = \sum_{w \in \{i, j, l\}} R_{w,t}\left(a_{w,t} \mid \mathbf{Z}_t^*\right).
\end{equation}

Since the scheduling and association decisions are fixed by Algorithm~1 and no inter-slot power-coupling constraint is imposed, the power allocation can be independently optimized for each time slot $t$. Accordingly, $\mathcal{P}_2$ is formulated as
\begin{subequations} \label{eq:problem_P2}
\begin{align}
\mathcal{P}_2: \quad & \max_{a_{w,t}} \sum_{w \in \{i, j, l\}} R_{w,t}\left(a_{w,t} \mid \mathbf{Z}_t^*\right) \\
&\text{s.t.} \quad \eqref{eq:const_power_sum}, \eqref{eq:const2}, \eqref{eq:const_aw_range}, \eqref{eq:const5}, \eqref{eq:const_rl_min}, \eqref{eq:const_sic}.
\end{align}
\end{subequations}

In summary, Algorithm~1 determines $z_{s,b,t}^*$ and $\beta_{s,b,m,t}^*$ from the slot-dependent large-scale gain matrix $\mathbf{G}_t$, avoiding repeated scheduling searches driven by small-scale fading. For each time slot, under the adopted repeated-scan greedy implementation, the selection of $N_B$ feasible beam-user pairs from $N_{\mathrm{tot}}M$ candidates requires a complexity of $\mathcal{O}(N_B N_{\mathrm{tot}}M)$, leading to $\mathcal{O}(TN_B N_{\mathrm{tot}}M)$ over the scheduling horizon. Algorithm~2 then searches $N_a$ candidate group-level power-allocation vectors for each scheduled NOMA group, resulting in a complexity of $\mathcal{O}(TN_a)$. Due to the sum-power constraint in \eqref{eq:const_power_sum}, only two of the three power-allocation coefficients are independent. Hence, for a uniform search step $\Delta$, $N_a = \mathcal{O}(1/\Delta^2)$, and the power-allocation complexity becomes $\mathcal{O}(T/\Delta^2)$. Therefore, the overall computational complexity of the designed LCD algorithm is $\mathcal{O}\left(T N_B N_{\mathrm{tot}}M + \frac{T}{\Delta^2}\right)$. For a fixed small $N_B$, the scheduling complexity reduces to $\mathcal{O}(TN_{\mathrm{tot}}M)$.

\section{PERFORMANCE ANALYSIS}\label{PERFORMANCE ANALYSIS}
In this section, we analyze the performance of the designed NOMA-CBH scheme in the considered multidimensional VLEO satellite communication system. The analysis is conducted on a per-snapshot basis under a two-layer decoupling that mirrors the designed LCD algorithm. Specifically, the large-scale-gain-driven scheduling in Algorithm~1 first determines the two co-scheduled ordinary users, denoted by $u_k$, together with their respective associated satellite-beam links $(s_k, b_k)$ for $k \in \{1, 2\}$. Let $g_k \triangleq |h_{s_k,b_k,u_k}|^2$ denote the corresponding instantaneous channel power gains before SIC ordering, with average channel power gains $\Omega_{u_k}$. Under the adopted Nakagami-$\mu$ fading model, $g_1$ and $g_2$ are modeled as independent but non-identically distributed Gamma random variables with rate parameters $\Lambda_k = \frac{\mu}{\Omega_{u_k}}$. Conditioned on the scheduling outcome, the SIC decoding order within the scheduled NOMA group is determined by the instantaneous channel power gains of these two co-scheduled users. Accordingly, the weak- and strong-user channel gains are respectively defined as $g_i = \min\{g_1, g_2\}$, and $g_j = \max\{g_1, g_2\}$. The GS is treated separately as a dedicated dual-satellite cooperative link, as discussed below Theorem~3. Based on this snapshot-wise analysis, the overall system performance is then evaluated by averaging the instantaneous probabilities over time to capture the dynamic characteristics of the VLEO constellation, with the accuracy of the analysis validated by Monte Carlo simulations in Section~V. For notational simplicity, the snapshot index $t$ is omitted in the following analytical expressions, while the large-scale channel parameters are understood to be conditioned on the current snapshot.

For the two pre-ordered ordinary-user links, the channel power gain $g_k$ follows a Gamma distribution with shape parameter $\mu$ and rate parameter $\Lambda_k = \mu/\Omega_{u_k}$. Hence, its cumulative distribution function (CDF) is denoted by $F_k(x)$. For the integer-valued Nakagami parameter $\mu$ adopted in the closed-form analysis, it can be expressed as $F_k(x) = 1 - e^{-\Lambda_k x} \sum_{n=0}^{\mu-1} \frac{(\Lambda_k x)^n}{n!}$. Since the two links generally have different average channel power gains, i.e., $\Omega_{u_1} \neq \Omega_{u_2}$, they are treated as independent but non-identically distributed rather than using the conventional i.i.d. order-statistics model. For the mandatorily included GS, the two coordinated satellite branches are treated separately from the ordinary-user ordering. Under the symmetric two-branch Nakagami-$\mu$ model adopted for the GS, $\Omega_l$ denotes the average channel power gain of each coordinated branch. Hence, the MRC-combined channel has shape parameter $2\mu$, while its equivalent rate parameter is $\Lambda_l = \frac{\mu}{\chi\Omega_l}$. For the analytical outage characterization in this section, the power-allocation vector is treated as a given feasible vector and remains fixed with respect to the instantaneous small-scale fading realizations. The same power-allocation setting is adopted in the Monte Carlo validation in Section V-B.

\subsection{The Outage Probability of $U_i$}
The outage of $U_i$ occurs when it fails to decode $x_i$, which can be expressed as the event that the received SINR at $U_i$ is below its target threshold $\gamma_i^{\mathrm{th}}$.
\begin{theorem} \label{therom1}
In the NOMA-CBH scheme, the closed-form expression for outage probability of $U_i$ is given by
\begin{equation}
\label{eq:out_Ui}
\scalebox{0.88}{%
$P_{U_i} = \begin{cases}
1, & P_i \le \gamma_i^{\text{th}}(P_j + P_l); \\[1.5ex]
\begin{aligned}
&1 - e^{-(\Lambda_1 + \Lambda_2)\xi_i} \\
&\quad \times \sum_{n=0}^{\mu-1} \sum_{r=0}^{\mu-1} \frac{(\Lambda_1 \xi_i)^n (\Lambda_2 \xi_i)^r}{n!r!},
\end{aligned} & P_i > \gamma_i^{\text{th}}(P_j + P_l)
\end{cases}$%
},
\end{equation}
\textit{where $\xi_i = \frac{\gamma_i^{\mathrm{th}}\sigma^2}{P_i - \gamma_i^{\mathrm{th}}(P_j + P_l)}$ represents the equivalent channel gain threshold for successful decoding at $U_i$.}
\end{theorem}

\textit{Proof: See Appendix A.}

\subsection{The Outage Probability of $U_j$}
The outage of $U_j$ occurs in the following two cases: i) $U_j$ fails to decode $x_i$, thereby failing the first step of the SIC process; and ii) $U_j$ successfully decodes $x_i$ but fails to decode $x_j$ due to insufficient SINR \cite{yue2024yiqutop}.

\begin{theorem} \label{therom2}
In the NOMA-CBH scheme, the closed-form expression for outage probability of $U_j$ can be expressed as
\begin{equation}
\scalebox{0.88}{$
P_{U_j} = \begin{cases}
1, & \begin{aligned} &P_i \le \gamma_i^{\mathrm{th}}(P_j + P_l) \\ &\text{or } P_j \le \gamma_j^{\mathrm{th}}P_l; \end{aligned} \\[16pt]
\begin{aligned}
&\left[ 1 - e^{-\Lambda_1 \xi_j} \sum_{n=0}^{\mu-1} \frac{(\Lambda_1 \xi_j)^n}{n!} \right] \\
&\times \left[ 1 - e^{-\Lambda_2 \xi_j} \sum_{r=0}^{\mu-1} \frac{(\Lambda_2 \xi_j)^r}{r!} \right],
\end{aligned} & \begin{aligned} &P_i > \gamma_i^{\mathrm{th}}(P_j + P_l) \\ &\text{and } P_j > \gamma_j^{\mathrm{th}}P_l \end{aligned}
\end{cases}
$},
\label{eq:out2,Kj=M}
\end{equation}
\textit{where $\xi_j = \max \left\{ \frac{\gamma_i^{\mathrm{th}}\sigma^2}{P_i - \gamma_i^{\mathrm{th}}(P_j + P_l)}, \frac{\gamma_j^{\mathrm{th}}\sigma^2}{P_j - \gamma_j^{\mathrm{th}} P_l} \right\}$ represents the equivalent channel-gain threshold for successful SIC decoding and subsequent decoding of $x_j$ at $U_j$.}
\end{theorem}

\textit{Proof: See Appendix B. }

\subsection{The Outage Probability of $U_l$}
The outage of $U_l$ occurs in the following three cases: i) $U_l$ fails to decode $x_i$, thereby failing the first step of the SIC process; ii) $U_l$ successfully decodes $x_i$ but fails to decode $x_j$ after the first SIC; and iii) $U_l$ successfully decodes both $x_i$ and $x_j$ but fails to decode $x_l$ due to insufficient SINR.

\begin{theorem} \label{theorem3}
In the NOMA-CBH system, the closed-form expression for outage probability of $U_l$ can be given by
\begin{equation}
\scalebox{0.98}{$
P_{U_l} = \begin{cases}
1, & \begin{aligned} &P_i \le \gamma_i^{\mathrm{th}}(P_j + P_l) \\ &\text{or } P_j \le \gamma_j^{\mathrm{th}}P_l; \end{aligned} \\[14pt]
1 - e^{-\Lambda_l \xi_l} \displaystyle\sum_{n=0}^{2\mu-1} \frac{(\Lambda_l \xi_l)^n}{n!}, & \begin{aligned} &P_i > \gamma_i^{\mathrm{th}}(P_j + P_l) \\ &\text{and } P_j > \gamma_j^{\mathrm{th}}P_l \end{aligned}
\end{cases}
$},
\label{eq:out3,Kl=M}
\end{equation}
where $\xi_l = \max \left\{ \frac{\gamma_i^{\mathrm{th}}\sigma^2}{P_i - \gamma_i^{\mathrm{th}}(P_j + P_l)}, \frac{\gamma_j^{\mathrm{th}}\sigma^2}{P_j - \gamma_j^{\mathrm{th}} P_l}, \frac{\gamma_l^{\mathrm{th}}\sigma^2}{P_l} \right\}$ represents the equivalent MRC channel-gain threshold required for successful successive decoding at $U_l$. Here, $P_l > 0$ follows from the feasible power-allocation assumption.
\end{theorem}

\textit{Proof: See Appendix C. }

\subsection{Diversity Order Analyses}
To gain deeper insight, the diversity order is evaluated as a crucial metric characterizing the slope of the outage probability curve at high signal-to-noise ratio (SNR). Formally, the diversity order is defined as
\begin{align}
d = - \lim_{\rho \to \infty} \frac{\log (P^{\infty}(\rho))}{\log \rho},
\label{eq:ddingyi}
\end{align}
where $P^{\infty}(\rho)$ denotes the asymptotic outage probability at high SNR. Here, $\rho = P_{\mathrm{tot}}/\sigma^2$, and the power-allocation coefficients are assumed to remain fixed and feasible as $\rho \to \infty$. Accordingly, $\xi_i$, $\xi_j$, and $\xi_l$ scale as $\mathcal{O}(\rho^{-1})$.

\begin{corollary}
Based on the analysis in \eqref{eq:out_Ui}, under the fixed feasible power-allocation setting, as $\rho \to \infty$, the asymptotic outage probability of $U_i$ in the NOMA-CBH scheme is given by
\begin{equation}
P_{U_i}^{\infty}
=
\frac{\Lambda_1^{\mu}+\Lambda_2^{\mu}}{\mu!}
\xi_i^{\mu}.
\label{eq:wuqiongout1}
\end{equation}
\end{corollary}

\begin{remark}
Since $\xi_i \propto \rho^{-1}$, substituting \eqref{eq:wuqiongout1} into \eqref{eq:ddingyi} yields the diversity order $d_i = \mu$. Hence, the non-identical average channel gains affect the asymptotic coefficient but do not change the diversity order of $U_i$.
\end{remark}

\begin{corollary}
Based on the analysis in \eqref{eq:out2,Kj=M}, under the fixed feasible power-allocation setting, as $\rho \to \infty$, the asymptotic outage probability of $U_j$ in the NOMA-CBH scheme is given by
\begin{equation}
P_{U_j}^{\infty}
=
\frac{(\Lambda_1\Lambda_2)^{\mu}}{(\mu!)^2}
\xi_j^{2\mu}.
\label{eq:wuqiongout2,K=M}
\end{equation}
\end{corollary}

\begin{remark}
Since $\xi_j \propto \rho^{-1}$, substituting \eqref{eq:wuqiongout2,K=M} into \eqref{eq:ddingyi} yields the diversity order $d_j = 2\mu$. This result shows that selecting the stronger user from the two independently faded scheduled links provides a diversity order of $2\mu$, even when the two links are non-identically distributed.
\end{remark}

\begin{corollary}
Based on the analysis in \eqref{eq:out3,Kl=M}, under the fixed feasible power-allocation setting, as $\rho \to \infty$, the asymptotic outage probability of $U_l$ in the NOMA-CBH scheme is given by
\begin{equation}
P_{U_l}^{\infty}
=
\frac{\Lambda_l^{2\mu}}{(2\mu)!}
\xi_l^{2\mu}.
\label{eq:50}
\end{equation}
\end{corollary}

\begin{remark}
Since $\xi_l \propto \rho^{-1}$, substituting \eqref{eq:50} into \eqref{eq:ddingyi} yields the diversity order $d_l = 2\mu$. This result reflects the diversity gain provided by the two independent satellite branches under MRC.
\end{remark}

\subsection{Throughput Analyses}
The satellite transmits information at a constant rate. Hence, the throughput of the NOMA-CBH scheme in delay-limited mode is defined as
\begin{align}
\bar{R}_w = (1 - P_{U_w}) R_w,
\label{tuntuliangdingyi}
\end{align}
where $R_w$ denotes the fixed target transmission rate of terminal $U_w$ in the delay-limited mode.

Combining~\eqref{tuntuliangdingyi} with the outage results derived in the previous subsections, using~\eqref{eq:out_Ui}, \eqref{eq:out2,Kj=M}, and~\eqref{eq:out3,Kl=M}, the system throughput of the NOMA-CBH scheme is given by
\begin{align}
    \bar{R}_{\mathrm{sum}} = (1 - P_{U_i})R_i + (1 - P_{U_j})R_j + (1 - P_{U_l})R_l.
    \label{tuntuliang}
\end{align}

\section{NUMERICAL RESULTS}\label{NUMERICAL RESULTS}
This section presents the numerical analysis of the designed NOMA-CBH scheme and the benchmark schemes. Section~V-A evaluates the optimization performance of the complete LCD algorithm. The VLEO coverage characteristics and user traffic-demand profiles are adopted from~\cite{chen2023vleo}, while the satellite-related parameters and beam-activation settings are configured according to 3GPP TR~38.811 and TR~38.821~\cite{3gpp_38811,3gpp_38821}. For notational clarity, the instantaneous physical-layer system sum rate in time slot $t$ is defined as $R_{\mathrm{sum},t} = \sum_{m\in\mathcal{M}}R_{m,t} + R_{l,t}$. Accordingly, the objective in~\eqref{eq:objective_func} maximizes the accumulated slot-wise sum-rate utility $\sum_{t\in\mathcal{T}}R_{\mathrm{sum},t}$. Since the number of time slots $T$ and the slot duration $T_{\mathrm{slot}}$ are fixed, maximizing this objective is equivalent, up to a positive constant factor, to maximizing the time-average system sum rate $\overline{R}_{\mathrm{sum}} = \frac{1}{T} \sum_{t\in\mathcal{T}} R_{\mathrm{sum},t}$, or the total transmitted information $B_{\mathrm{tot}} = T_{\mathrm{slot}} \sum_{t\in\mathcal{T}} R_{\mathrm{sum},t}$. Accordingly, the rate quantities reported in bps in the numerical results correspond to time-averaged rates, whereas $B_{\mathrm{tot}}$ is measured in bits. For the system-level evaluation, the performance metrics are evaluated through at least 20 independent Monte Carlo simulations, each spanning $T=24$ time slots.

Each Monte Carlo realization spans $T=24$ time slots with $T_{\mathrm{slot}}=0.1$ s, corresponding to a total observation interval of $2.4$ s. For the slot-wise geometric evolution, we adopt a simplified constant-altitude circular-orbit kinematic model rather than a full orbital-element propagation. The Earth is approximated as a sphere of radius $R_E$, with orbital radius $r_{\mathrm{orb}}=R_E+h_{\mathrm{alt}}$ and angular velocity $\omega=v_{\mathrm{sat}}/r_{\mathrm{orb}}$. The initial geometric state of satellite $s$ is represented by its Earth-centered central angle $\vartheta_{s,0}$, which is randomly generated within $[-5^\circ,5^\circ]$ and retained only when the minimum-elevation requirement is satisfied over the simulated horizon. The corresponding slot-dependent central angle is updated as
\begin{equation}
\vartheta_{s,t} = \vartheta_{s,0} + (t-1)\omega T_{\mathrm{slot}}.
\label{eq:orbital_evolution}
\end{equation}

The ground terminals are treated as fixed over each realization, and the short-term satellite--ground geometry is represented using the above Earth-centered spherical model rather than a full ECI/ECEF state propagation. Owing to the short $2.4$-s observation interval, Earth rotation, atmospheric-drag-induced orbital decay, and higher-order orbital perturbations are neglected in the trajectory propagation.

In Section~V-B, Monte Carlo simulations validate the analytical results using fixed feasible power-allocation coefficients and an independently swept normalized SNR $\rho=P_{\mathrm{tot}}/\sigma^2$; Algorithm~2 is not re-optimized. The VLEO large-scale channel gains are normalized by their ensemble-average values, with reference average gains of one for the ordinary-user links and two for each GS branch. Hence, $\rho$ in Figs.~7 and~8 denotes a normalized link-level SNR for analytical verification rather than the absolute received SNR obtained from the unnormalized link budget. The remaining parameters are summarized in Table~I.

\begin{table}[htbp]
\caption{Simulation parameters}
\label{tab:sim_params}
\centering
\setlength{\arrayrulewidth}{0.5pt}
\renewcommand{\arraystretch}{1.3}
\begin{tabular}{|c|c|}
\hline
\textbf{Parameter} & \textbf{Value} \\ \hline
Bandwidth, $B_{\max}$ & 400 MHz \\ \hline
Center frequency, $f_c$ & 11.7 GHz \\ \hline
Fixed power-allocation coefficients, $[a_i, a_j, a_l]$ & $[0.70, 0.25, 0.05]$ \\ \hline
Minimum rate requirement, $R_m^{\min}$ & 1 BPCU \\ \hline
Minimum elevation angle, $\theta_{\min}$ & $20^\circ$ \\ \hline
MRC combining-loss factor, $\chi$ & 0.8 \\ \hline
Noise Power, $\sigma^2$ & -113 dBW \\ \hline
Nakagami- $\mu$ parameter, $\mu$ & 2 \\ \hline
Number of time slots, $T$ & 24 \\ \hline
Number of active beams, $N_B$ & 2 \\ \hline
Number of VLEO satellites, $S$ & 6 \\ \hline
Number of candidate ground users, $M$ & 4 \\ \hline
Number of power-allocation candidates, $N_a$ & 33 \\ \hline
Power-allocation search step size, $\Delta$ & 0.05 \\ \hline
Standard deviation of 2D attitude jitter, $\sigma_p$ & $0.1^{\circ}$ \\ \hline
Time-slot duration, $T_{\text{slot}}$ & 0.1 s \\ \hline
Transmit power budget per active beam, $P_{\mathrm{tot}}$ & 25 dBW \\ \hline
Total number of candidate beams, $N_{\mathrm{tot}}$ & 24 \\ \hline
VLEO satellite orbital speed, $v_{\mathrm{sat}}$ & 7.70 km/s \\ \hline
VLEO satellite altitude, $h_{\mathrm{alt}}$ & 360 km \\ \hline
Satellite covers the ground longitude range & $[100^{\circ} E , 105^{\circ} E]$ \\ \hline
Satellite covers the ground latitude range & $[-1.5^{\circ} S , 1.5^{\circ} N]$ \\ \hline
Satellite transmit antenna gain, $G_t$ & 35 dBi \\ \hline
User receive antenna gain, $G_r$ & 5 dBi \\ \hline
GS receive antenna gain, $G_l$ & 53.7 dBi \\ \hline
Residual Doppler factor & 0.05 \\ \hline
Effective coherence time, $T_{\mathrm{eff}}$ & 12 $\mu$s \\ \hline
Zenith atmospheric attenuation, $L_{\mathrm{zen}}$ & 3 dB \\ \hline
\end{tabular}
\end{table}

\subsection{Simulation Results of Optimization}
In this subsection, the performance of the designed NOMA-CBH scheme is evaluated against three conventional benchmark schemes and an exhaustive joint benchmark:
\begin{enumerate}
    \item[1)] \textit{Orthogonal Multiple Access with Coordinated Beam Hopping (OMA-CBH):} The OMA-CBH scheme serves as a baseline where orthogonal resources are assigned to users within the activated beams. It corresponds to a simplified version of $\mathcal{P}_0$ that focuses solely on $z_{s,b,t}$ and ensures that only one user is served per resource block. In particular, the OMA-CBH scheme adopts a time division multiple access transmission strategy, in which the three served terminals are allocated equal orthogonal time fractions within each transmission slot.

    \item[2)] \textit{Random User Matching NOMA with Coordinated Beam Hopping (Random-NOMA-CBH):} This benchmark replaces the large-scale-gain-based beam-user scheduling of Algorithm~1 with random beam-user selection. Specifically, $N_B$ active beams are randomly selected from the candidate beam pool, and two ground users are randomly chosen from the candidate user pool, while the GS is mandatorily included~\cite{zheng2024random}. Once this topology is fixed, $a_{w,t}$ is optimized using the same discrete search as in Algorithm~2 to maximize the sum rate.

    \item[3)] \textit{Fixed Power Allocation NOMA with Coordinated Beam Hopping (F-NOMA-CBH):} This scheme retains the same large-scale-gain-based greedy beam-user scheduling as Algorithm~1, while simplifying the power-domain allocation~\cite{deng2024fnoma}. Specifically, $z_{s,b,t}$ and $\beta_{s,b,m,t}$ are determined according to the large-scale-gain criterion described in Algorithm~1; however, $a_{w,t}$ is fixed at empirical values rather than being dynamically updated through the discrete search in Algorithm~2.

    \item[4)] \textit{Exhaustive Joint Benchmark:} This benchmark jointly enumerates all feasible beam-user scheduling configurations and the $N_a$ candidate power-allocation vectors, and selects the combination yielding the maximum system sum rate. It serves as an optimality reference for evaluating the performance gap and computational complexity of the proposed LCD algorithm.
\end{enumerate}

\begin{figure}[t!]
    \begin{center}
        \includegraphics[width=3.8in]{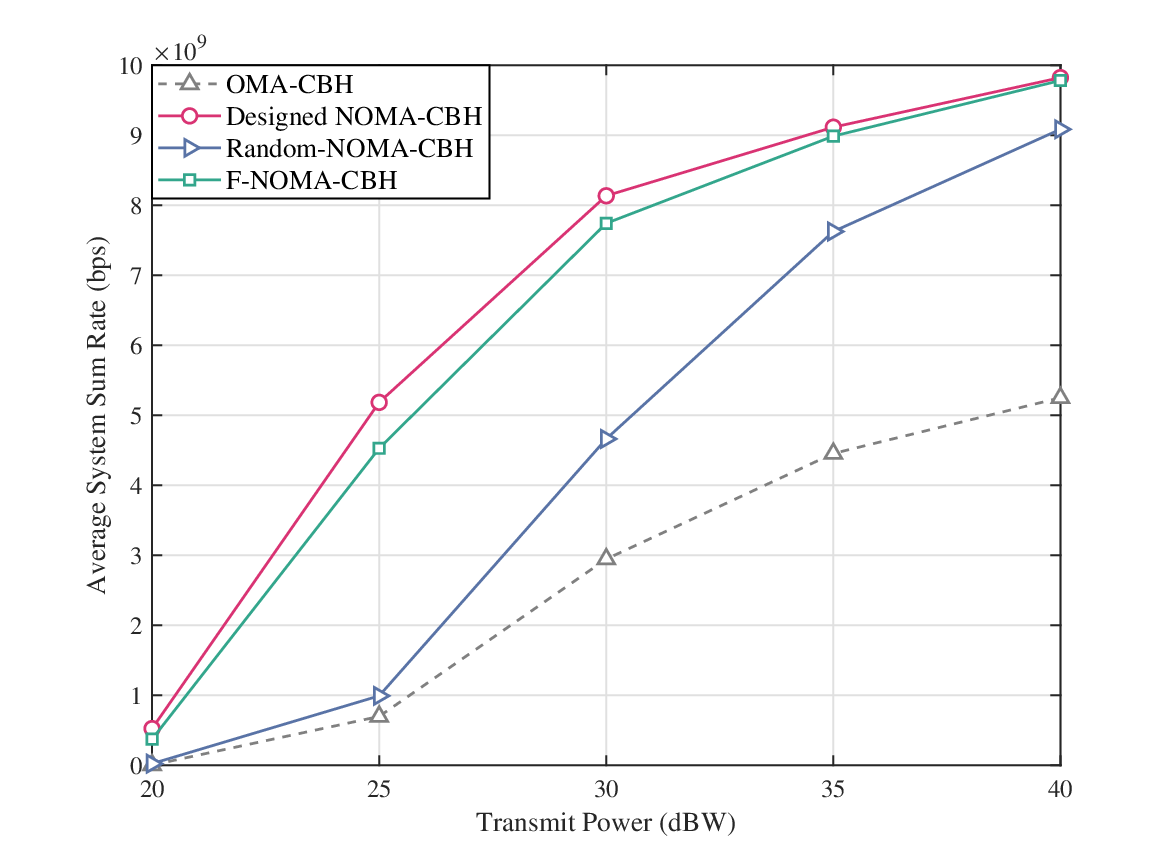}
        \caption{System sum rate versus transmit power comparing the NOMA-CBH scheme with benchmark schemes.}
        \label{Fig. 2}
    \end{center}
\end{figure}
Fig.~2 illustrates the system sum rate as a function of transmit power for the designed NOMA-CBH scheme and comparative benchmark methods. The NOMA-CBH scheme consistently achieves the highest sum rate, and the gap with respect to OMA-CBH widens as the transmit power increases. This performance gain is attributed to the combined effect of large-scale-gain-based greedy beam-user scheduling and adaptive power allocation, which effectively mitigates inter-user interference. In contrast, F-NOMA-CBH is constrained by static power coefficients, whereas Random-NOMA-CBH suffers from suboptimal topology selection. The distinct performance gaps they exhibit isolate and corroborate the individual contributions of the two cascaded stages, thereby validating the LCD decoupling rationale established in Section~III from an algorithmic perspective. These results confirm the spectral efficiency advantage of the designed NOMA-CBH scheme across varying power budgets.

\begin{figure}[t!]
    \centering
    \subfigure[Optimality-Gap Performance Evaluation.]{
        \includegraphics[width=3.8in]{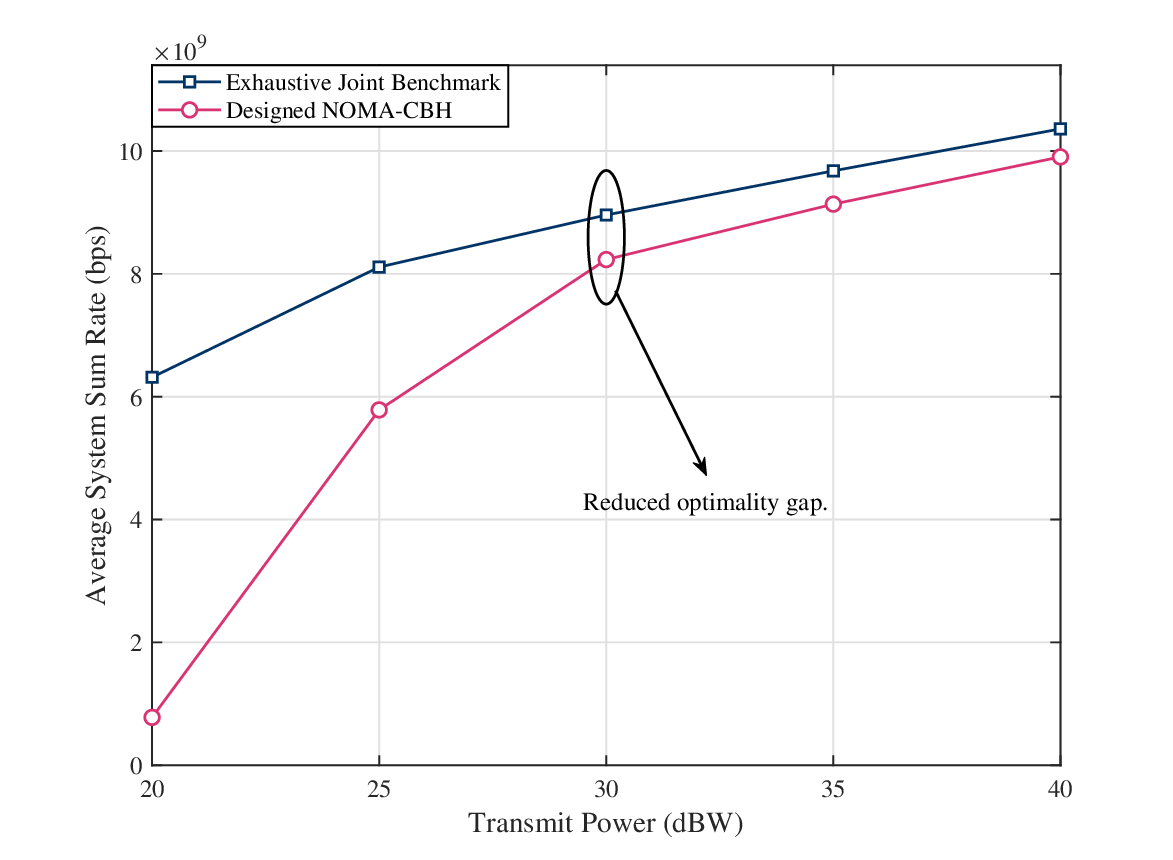}
        \label{Fig.3a}
    }
    
    \subfigure[Computational-Runtime Scalability.]{
        \includegraphics[width=3.8in]{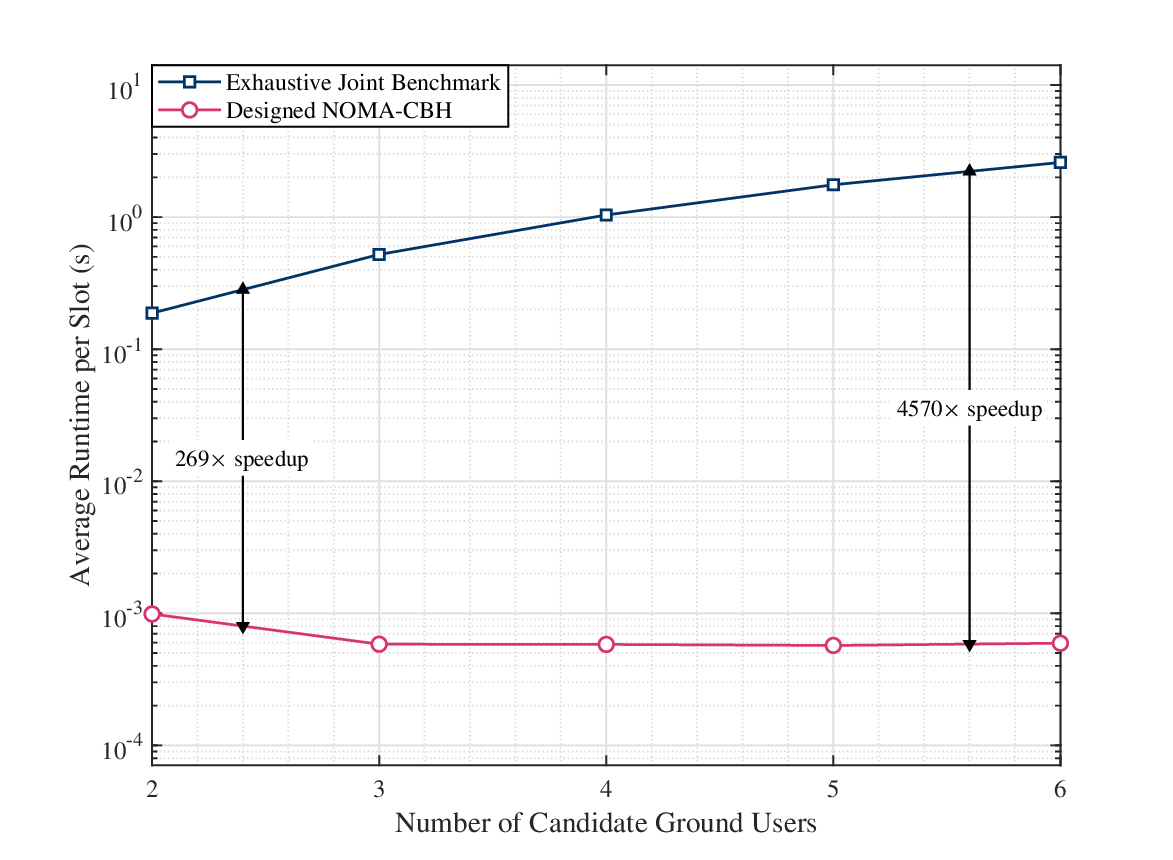}
        \label{Fig.3b}
    }
    \caption{Optimality-gap and computational-runtime comparison between the designed NOMA-CBH scheme and the exhaustive joint benchmark.}
    \label{Fig.3}
\end{figure}
Fig.~3 evaluates the performance and computational efficiency of the designed NOMA-CBH scheme against the exhaustive joint benchmark. As shown in Fig.~3(a), the performance gap is relatively pronounced in the low-power regime, since the large-scale-gain-based beam-user scheduling is decoupled from the subsequent QoS- and SIC-constrained power allocation. As the transmit power increases, the feasibility of the scheduled NOMA groups improves, and the designed scheme gradually approaches the exhaustive benchmark. In addition, Fig.~3(b) shows that the computational runtime of the exhaustive benchmark increases rapidly with the number of candidate ground users due to the joint enumeration of feasible scheduling configurations and the 33 candidate power-allocation vectors. In contrast, the proposed LCD algorithm only performs greedy beam-user matching followed by power allocation for the resulting scheduled group, leading to significantly lower computational complexity. Specifically, as the number of candidate ground users $M$ scales from 2 to 6, the runtime speedup of the proposed LCD algorithm over the exhaustive benchmark increases from approximately 269$\times$ to 4570$\times$. These results demonstrate that the proposed LCD algorithm achieves a favorable tradeoff between system performance and computational efficiency.

For the traffic-demand evaluations in Figs.~4 and~5, let $D^{\mathrm{agg}}$ denote the aggregate requested traffic load for the considered service scenario or traffic-demand operating point. The current LCD framework is capacity-oriented rather than queue-aware; hence, $D^{\mathrm{agg}}$ is treated as an exogenous evaluation quantity and does not enter the beam-user scheduling or power-allocation decisions in Algorithms~1 and~2. For the $n$th Monte Carlo realization, after the physical-layer transmission rates are determined in time slot $t$, the instantaneous aggregate achievable rate is given by $R_{\mathrm{sum},t}^{(n)} = R_{i,t}^{(n)} + R_{j,t}^{(n)} + R_{l,t}^{(n)}$. Accordingly, the aggregate delivered traffic in time slot $t$ is defined as $R_{\mathrm{del},t}^{\mathrm{agg},(n)} = \min \left\{ D^{\mathrm{agg}}, R_{\mathrm{sum},t}^{(n)} \right\}$. The reported average aggregate delivered traffic is therefore calculated as $\overline{R}_{\mathrm{del}}^{\mathrm{agg}} = \frac{1}{N_{\mathrm{MC}}T} \sum_{n=1}^{N_{\mathrm{MC}}} \sum_{t\in\mathcal{T}} R_{\mathrm{del},t}^{\mathrm{agg},(n)}$, where $N_{\mathrm{MC}}$ denotes the number of independent Monte Carlo realizations. If no candidate power-allocation vector satisfies the QoS and SIC constraints for the scheduled NOMA group in a given slot, the corresponding aggregate achievable rate is set to zero and the slot is retained in the averaging. Therefore, the requested traffic load limits the useful delivered service after the physical-layer capacity has been determined, while it does not alter the scheduling or power-allocation decisions of the current capacity-oriented LCD framework.

\begin{figure}[t!]
    \begin{center}
        \includegraphics[width=3.8in]{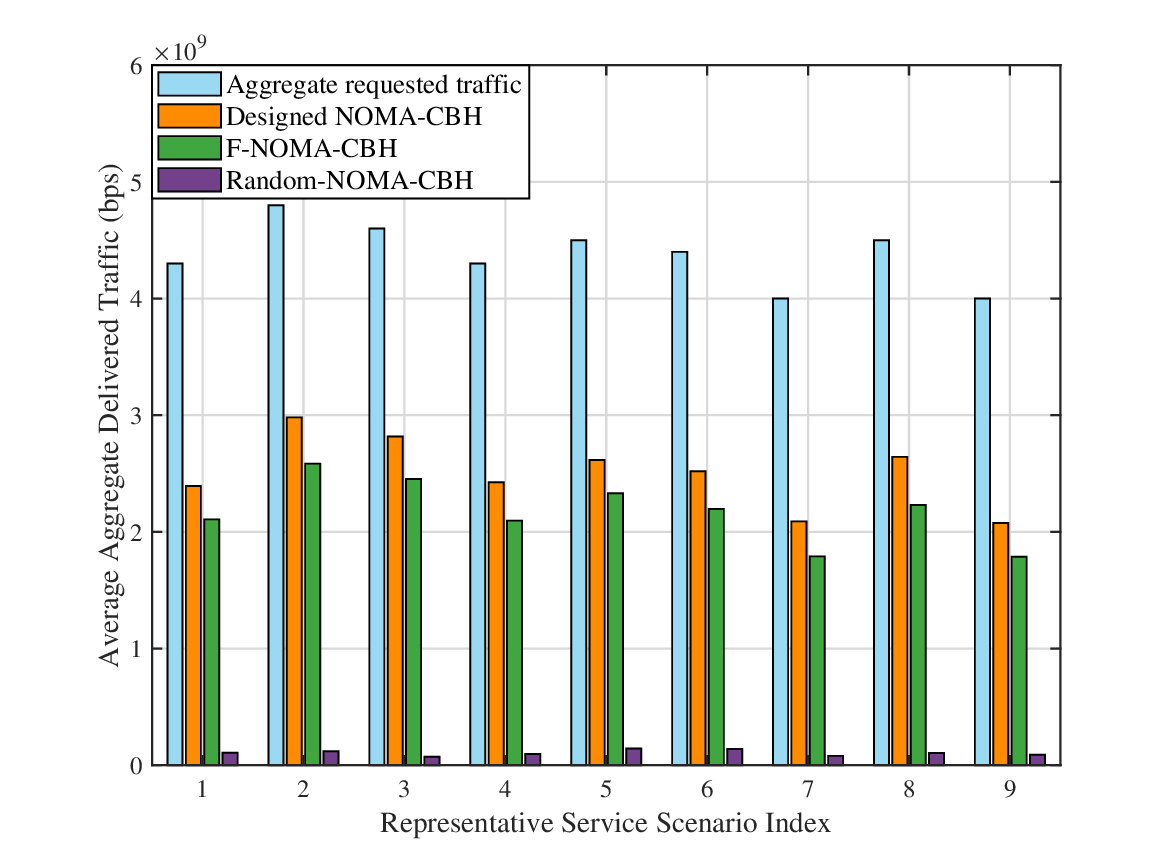}
        \caption{Average aggregate delivered traffic across representative service scenarios for the designed NOMA-CBH scheme and benchmark schemes.}
        \label{Fig. 4}
    \end{center}
\end{figure}
Fig.~4 compares the average aggregate delivered traffic across nine representative spatial service scenarios under heterogeneous aggregate requested traffic loads. The nine indices correspond to nine independently evaluated spatial scenarios generated with different angular offsets, while each scenario adopts the same two-beam NOMA-CBH configuration. It is observed that the designed NOMA-CBH scheme consistently achieves the highest aggregate delivered traffic among the considered schemes. Averaged over the nine representative scenarios, the designed NOMA-CBH scheme achieves approximately $2.51$~Gbps, compared with approximately $2.18$~Gbps for F-NOMA-CBH and $0.11$~Gbps for Random-NOMA-CBH. The remaining gap between the requested and delivered traffic mainly results from slots in which no candidate power-allocation vector satisfies the QoS and SIC constraints; such slots are conservatively counted as zero-service slots in the averaging. These results demonstrate the advantage of jointly adapting beam-user matching and power allocation under heterogeneous service conditions.

\begin{figure}[t!]
    \begin{center}
        \includegraphics[width=3.8in]{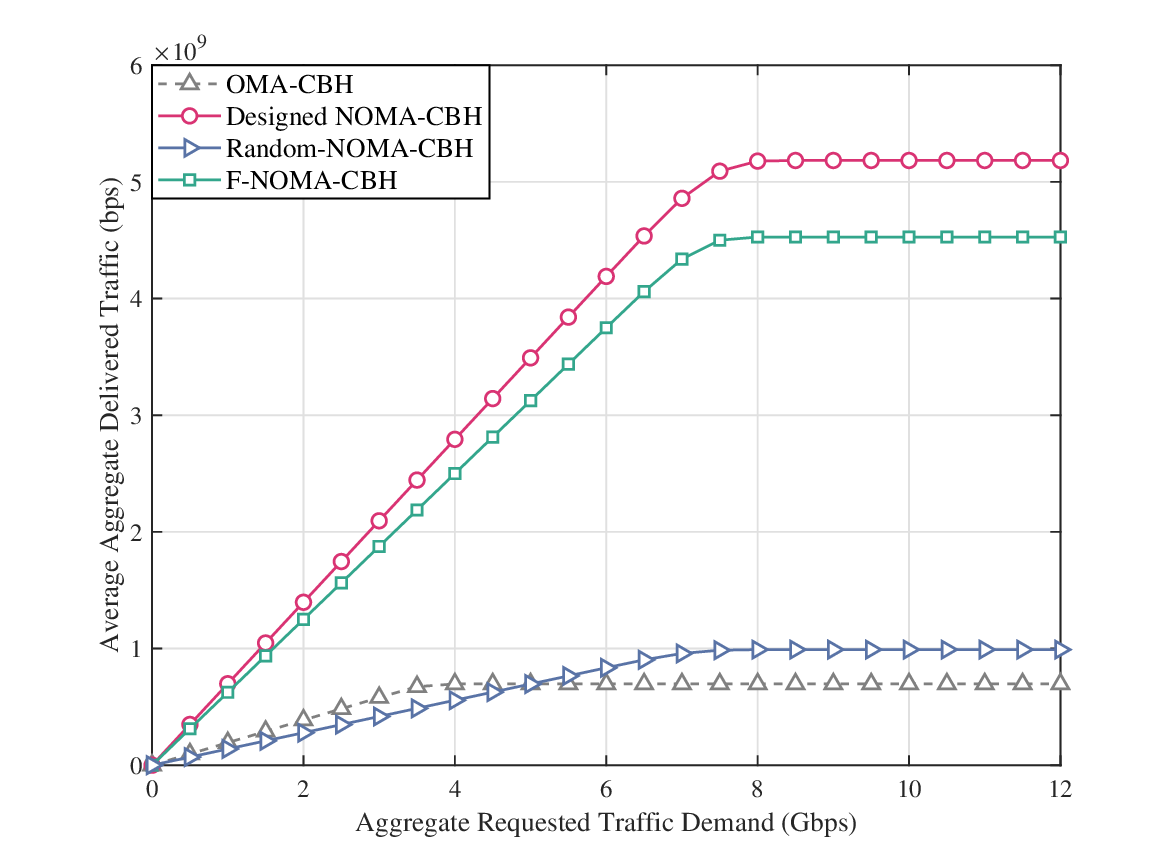}
        \caption{Average aggregate delivered traffic versus aggregate requested traffic demand for the designed NOMA-CBH scheme and benchmark schemes.}
        \label{Fig. 5}
    \end{center}
\end{figure}

Fig.~5 illustrates the average aggregate delivered traffic versus the aggregate requested traffic demand at a fixed transmit power of $25$~dBW. In the low-demand regime, the delivered traffic increases approximately linearly with the requested load, while the different slopes reflect the fractions of serviceable slots under the corresponding QoS and SIC constraints. As the requested traffic increases, the curves gradually approach their respective physical-layer capacity limits. In particular, the designed NOMA-CBH scheme reaches the highest saturation level of approximately $5.18$~Gbps, followed by F-NOMA-CBH at approximately $4.53$~Gbps, Random-NOMA-CBH at approximately $0.99$~Gbps, and OMA-CBH at approximately $0.70$~Gbps. This behavior indicates that the designed scheme provides a higher aggregate service capability over a broad traffic-load range, while the high-demand regime is ultimately limited by the achievable physical-layer capacity.

\begin{figure}[t!]
    \begin{center}
        \includegraphics[width=3.8in]{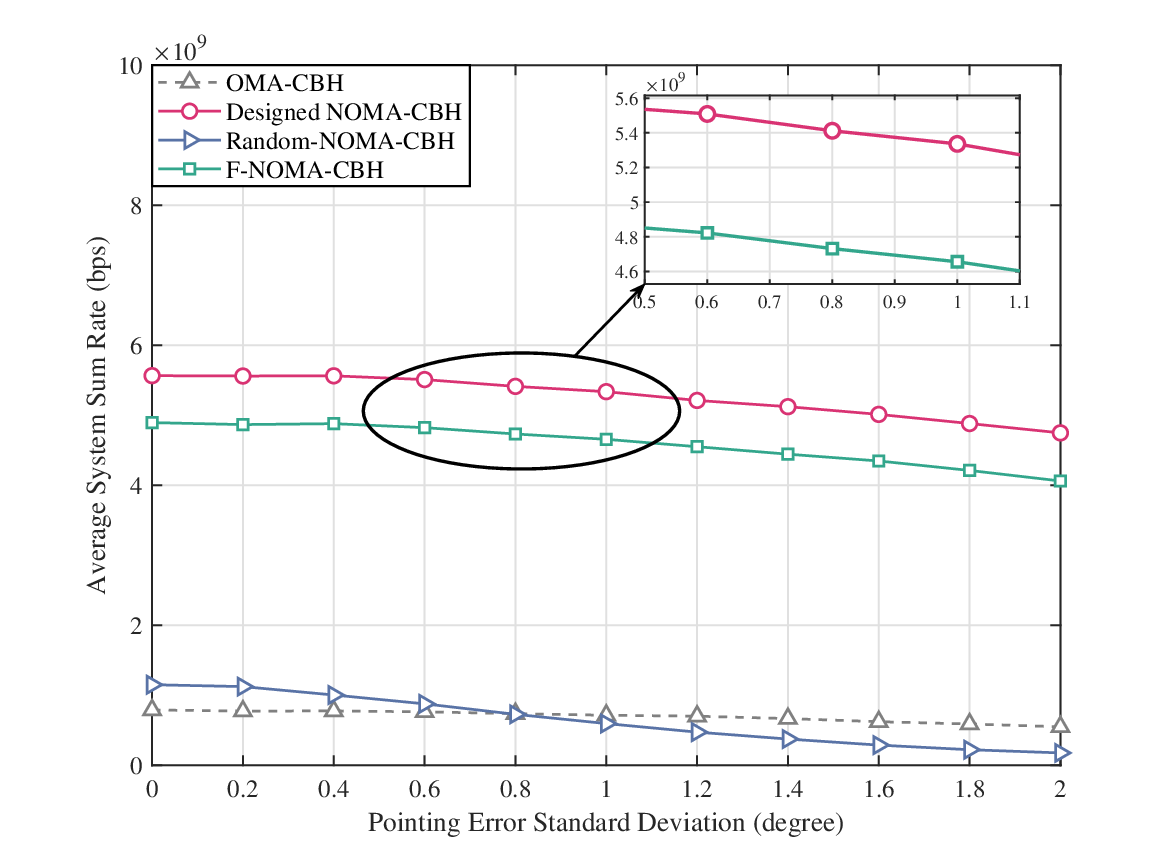}
        \caption{System sum rate under different pointing error standard deviations for NOMA-CBH and benchmark schemes.}
        \label{Fig. 6}
    \end{center}
\end{figure}
Fig.~6 evaluates the robustness of the considered schemes against satellite pointing errors. It is observed that the designed NOMA-CBH scheme maintains a relatively stable sum rate even when the standard deviation of the pointing jitter increases to $2^{\circ}$, demonstrating strong robustness against attitude perturbations. Although the NOMA-CBH schemes experience performance degradation as the pointing error grows, the reduction of the designed scheme remains well controlled owing to the joint adaptation of beam-user matching and power allocation. In comparison, F-NOMA-CBH suffers a larger performance loss due to its fixed power-allocation strategy, while Random-NOMA-CBH is more sensitive to pointing-induced channel degradation because of its random beam-user matching. Notably, when the pointing error becomes sufficiently large, Random-NOMA-CBH performs even worse than OMA-CBH. This is because random beam-user matching may select severely degraded links under large pointing perturbations, thereby increasing the likelihood that the coupled SIC and QoS requirements of NOMA cannot be simultaneously satisfied, whereas OMA avoids intra-group interference and SIC constraints and is therefore less vulnerable in such unfavorable channel conditions. These results validate the robustness of the designed NOMA-CBH scheme against pointing-induced channel impairments and highlight the importance of jointly adapting user matching and power allocation to dynamic channel variations.

\subsection{Simulation Results of Performance Analysis}
\begin{figure}[t!]
    \begin{center}
        \includegraphics[width=3.8in]{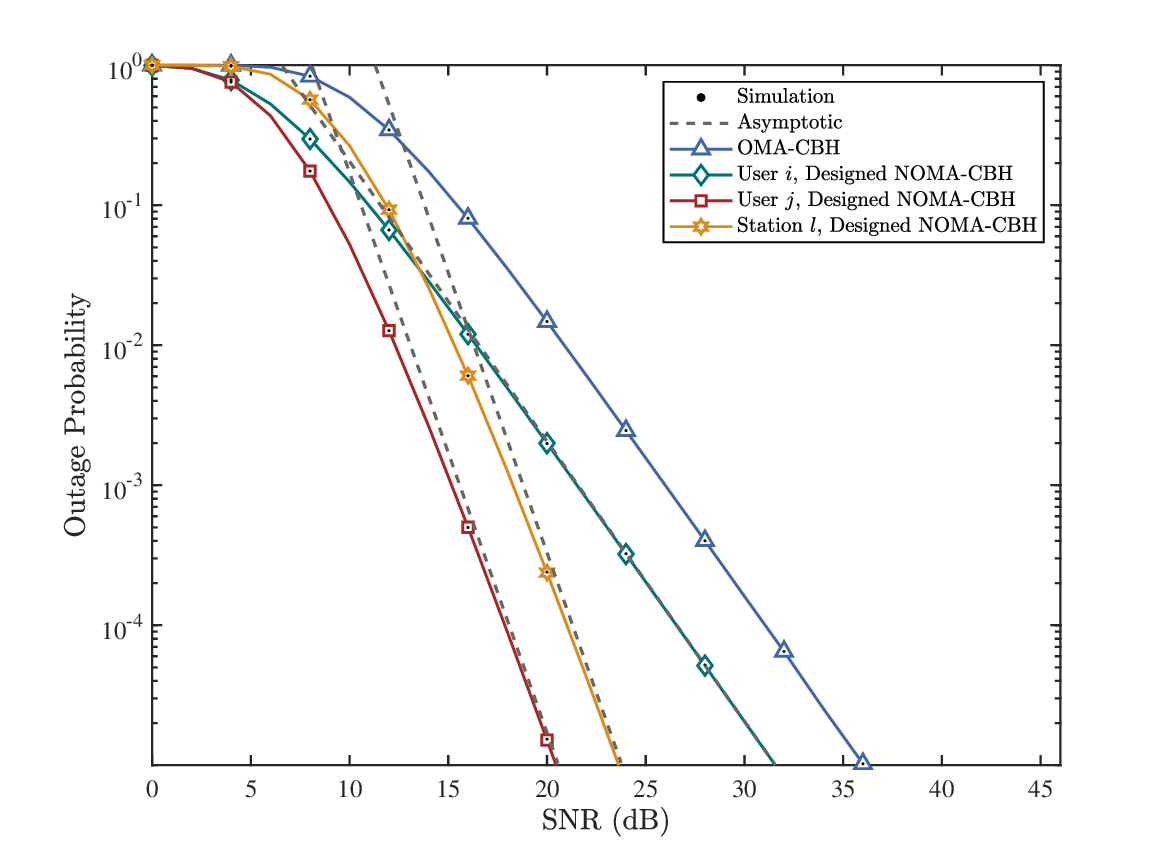}
        \caption{Outage probability of NOMA-CBH scheme versus SNR, with $R_i = R_j = R_l = 1$ BPCU.}
        \label{Fig.7}
    \end{center}
\end{figure}
Fig. 7 illustrates the outage probability versus SNR and evaluates the reliability of $U_i$, $U_j$, and $U_l$ under the NOMA-CBH scheme. As SNR increases, the outage probability decreases monotonically without an observable error floor, indicating effective interference suppression. In the high-SNR regime, the outage curves of the ground users follow their respective asymptotic slopes, which is in line with the discussion in \textbf{Remark 1} and \textbf{Remark 2}. Moreover, owing to the coordinated dual-satellite transmission, the GS achieves an additional diversity gain, which verifies the conclusion in \textbf{Remark 3}. This corroborates that, by leveraging coordinated dual-satellite transmission, the NOMA-CBH scheme substantially enhances link reliability in highly dynamic VLEO satellite systems. Taken together, these results validate the analytical outage characterization and demonstrate the reliability advantage of the designed NOMA-CBH scheme over the considered OMA-CBH reference.

\begin{figure}[t!]
    \begin{center}
        \includegraphics[width=3.8in]{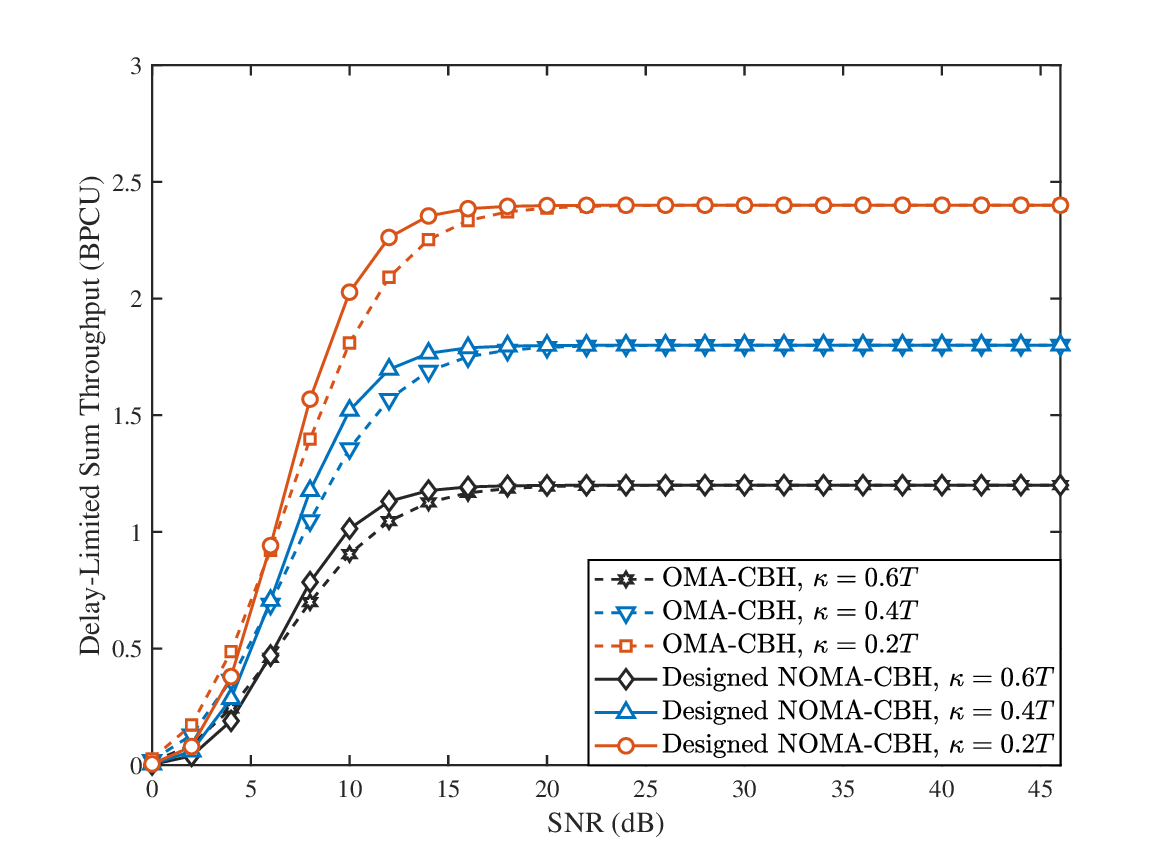}
        \caption{Impact of resource constraints on system throughput versus SNR.}
        \label{Fig. 8}
    \end{center}
\end{figure}
Fig. 8 plots the delay-limited system throughput versus SNR under different resource-availability constraints. Let $\kappa$ denote the number of constrained slots within the observation window of $T$ slots. These constrained slots are assumed to provide no transmission opportunity; hence, only the fraction $1-\kappa/T$ of the observation window is available for useful transmission. Accordingly, the constrained delay-limited throughput is evaluated as $\bar{R}_{\mathrm{sum}}^{(\kappa)} = (1-\kappa/T)\bar{R}_{\mathrm{sum}}$, where $\bar{R}_{\mathrm{sum}}$ is given by~\eqref{tuntuliang}. This resource-availability factor is introduced only in the conditional delay-limited throughput evaluation of Section~V-B and does not modify the beam-user scheduling problem in Section~III. It can be observed that the system throughput first increases with SNR and then approaches a resource-availability-limited ceiling. Although the NOMA-CBH and OMA-CBH schemes approach the same ceiling under a given $\kappa$ at sufficiently high SNR, the NOMA-CBH scheme reaches this ceiling at a substantially lower SNR, demonstrating substantial power savings. Moreover, increasing $\kappa$ from $0.2T$ to $0.6T$ reduces the available transmission fraction and therefore lowers the corresponding throughput ceiling.

\section{Conclusion}
In this paper, we have investigated the NOMA-CBH scheme for VLEO satellite communication systems. A joint resource optimization problem has been formulated by coupling beam scheduling and power allocation, with the objective of maximizing the system sum rate. To address the coupling between beam scheduling and power allocation, the LCD algorithm has been designed to sequentially determine the beam scheduling and power allocation. Furthermore, we have derived the closed-form expressions for outage probabilities of ground users and GS, followed by asymptotic analysis to characterize the corresponding diversity orders. Moreover, the system throughput under delay-limited transmission has been evaluated. Numerical results have verified the analytical derivations and demonstrated that the designed NOMA-CBH scheme provides substantial performance improvements over the considered benchmark schemes. Incorporating practical implementation impairments, such as non-coordinated cochannel inter-beam interference and control-link synchronization overhead, into large-scale VLEO coordinated scheduling constitutes an important direction in our future work.

\appendices
\section*{Appendix~A: Proof of Theorem \ref{therom1}} \label{Appendix:A}
\renewcommand{\theequation}{A.\arabic{equation}}
\setcounter{equation}{0}

\textit{The receiver-specific outage probability of $U_i$ can be expressed as}
\begin{align}
    P_{U_i} = \underbrace{\Pr\left(\gamma_i < \gamma_i^{\mathrm{th}}\right)}_{J_{1}}.
    \label{eq:A1}
\end{align}

\textit{According to \eqref{izhongduan} and the instantaneous ordering $g_i=\min\{g_1,g_2\}$, $J_{1}$ can be derived as}
\begin{equation}
\begin{aligned}[b]
    J_{1} &= \Pr\left( \frac{g_i P_i}{g_i(P_j+P_l)+\sigma^2} < \gamma_i^{\mathrm{th}} \right) \\
    &= \Pr(g_i < \xi_i) \\
    &= F_{g_i}(\xi_i),
\end{aligned}
\label{eq:A2} 
\end{equation}
\textit{where $\xi_i = \frac{\gamma_i^{\mathrm{th}}\sigma^2}{P_i-\gamma_i^{\mathrm{th}}(P_j+P_l)}$ for $P_i > \gamma_i^{\mathrm{th}}(P_j+P_l)$. If $P_i \le \gamma_i^{\mathrm{th}}(P_j+P_l)$, the target SINR cannot be achieved for any finite $g_i$, and thus $P_{U_i}=1$.}

\textit{Based on \eqref{eq:A2}, since $g_1$ and $g_2$ are independent but non-identically distributed, $F_{g_i}(\xi_i)$ can be calculated as}
\begin{align}
    F_{g_i}(\xi_i) &= 1-\Pr(g_1>\xi_i,g_2>\xi_i) \notag\\
    &= 1-\left[1-F_1(\xi_i)\right] \left[1-F_2(\xi_i)\right].
    \label{eq:A3}
\end{align}

\textit{Using $\Lambda_k=\frac{\mu}{\Omega_{u_k}}$ and the complementary CDF of $g_k$ for integer-valued Nakagami-$\mu$ fading, given by $1-F_k(\xi_i) = e^{-\Lambda_k \xi_i} \sum_{n=0}^{\mu-1} \frac{(\Lambda_k \xi_i)^n}{n!}$ for $k\in\{1,2\}$, $J_{1}$ can be calculated as}
\begin{equation}
\scalebox{0.89}{$
\begin{aligned}[b]
    J_{1} &= 1 - \left( e^{-\Lambda_1 \xi_i} \sum_{n=0}^{\mu-1} \frac{(\Lambda_1 \xi_i)^n}{n!} \right) \left( e^{-\Lambda_2 \xi_i} \sum_{r=0}^{\mu-1} \frac{(\Lambda_2 \xi_i)^r}{r!} \right).
\end{aligned}
$}
\label{eq:A4}
\end{equation}

\textit{Furthermore, the expression can be simplified to}
\begin{align}
    J_{1} = 1 - e^{-(\Lambda_1+\Lambda_2)\xi_i} \sum_{n=0}^{\mu-1} \sum_{r=0}^{\mu-1} \frac{ (\Lambda_1 \xi_i)^n (\Lambda_2 \xi_i)^r }{ n!r! }.
    \label{eq:A5}
\end{align}

\textit{Substituting \eqref{eq:A5} into \eqref{eq:A1}, together with the infeasible case $P_i \le \gamma_i^{\mathrm{th}}(P_j+P_l)$, we can obtain \eqref{eq:out_Ui}. The proof is completed.}

\section*{Appendix~B: Proof of Theorem \ref{therom2}} \label{Appendix:B}
\renewcommand{\theequation}{B.\arabic{equation}}
\setcounter{equation}{0}

\textit{The receiver-specific outage probability of $U_j$ can be expressed as}
\begin{align}
    P_{U_j} = \underbrace{1 - \Pr \left( \gamma_{j \to i} \ge \gamma_i^{\mathrm{th}}, \gamma_j \ge \gamma_j^{\mathrm{th}} \right)}_{J_{2}}.
    \label{eq:B1}
\end{align}

\textit{By substituting \eqref{jjiancei} and \eqref{jjiancezishen} into \eqref{eq:B1}, and using the instantaneous ordering $g_j = \max\{g_1, g_2\}$, the expression for $J_{2}$ is formulated as}
\begin{align}
    J_{2} &= 1 - \Pr \left( g_j \ge \frac{\gamma_i^{\mathrm{th}}\sigma^2}{P_i - \gamma_i^{\mathrm{th}}(P_j + P_l)}, g_j \ge \frac{\gamma_j^{\mathrm{th}}\sigma^2}{P_j - \gamma_j^{\mathrm{th}}P_l} \right) \notag \\
    &= \Pr(g_j < \xi_j) \notag \\
    &= F_{g_j}(\xi_j), \label{eq:B2}
\end{align}
\textit{where $\xi_j = \max \left\{ \frac{\gamma_i^{\mathrm{th}}\sigma^2}{P_i - \gamma_i^{\mathrm{th}}(P_j + P_l)}, \frac{\gamma_j^{\mathrm{th}}\sigma^2}{P_j - \gamma_j^{\mathrm{th}}P_l} \right\}$ for $P_i > \gamma_i^{\mathrm{th}}(P_j + P_l)$ and $P_j > \gamma_j^{\mathrm{th}}P_l$. If either $P_i \le \gamma_i^{\mathrm{th}}(P_j + P_l)$ or $P_j \le \gamma_j^{\mathrm{th}}P_l$, at least one required decoding step is infeasible for any finite $g_j$, and thus $P_{U_j} = 1$.}

\textit{Based on \eqref{eq:B2}, since $g_1$ and $g_2$ are independent but non-identically distributed, $F_{g_j}(\xi_j)$ can be calculated as}
\begin{align}
    F_{g_j}(\xi_j) &= \Pr(g_1 < \xi_j, g_2 < \xi_j) \notag\\
    &= F_1(\xi_j)F_2(\xi_j).
    \label{eq:B3}
\end{align}

\textit{Using $\Lambda_k = \frac{\mu}{\Omega_{u_k}}$ and the CDF of $g_k$ for integer-valued Nakagami-$\mu$ fading, given by $F_k(\xi_j) = 1 - e^{-\Lambda_k \xi_j} \sum_{n=0}^{\mu-1} \frac{(\Lambda_k \xi_j)^n}{n!}$ for $k \in \{1, 2\}$, $J_{2}$ is given by}
\begin{equation}
\scalebox{0.85}{$\displaystyle
    J_{2} = \left( 1 - e^{-\Lambda_1 \xi_j} \sum_{n=0}^{\mu-1} \frac{(\Lambda_1 \xi_j)^n}{n!} \right) \left( 1 - e^{-\Lambda_2 \xi_j} \sum_{r=0}^{\mu-1} \frac{(\Lambda_2 \xi_j)^r}{r!} \right).
$}
\label{eq:B4}
\end{equation}

\textit{Substituting \eqref{eq:B4} into \eqref{eq:B1}, together with the infeasible cases $P_i \le \gamma_i^{\mathrm{th}}(P_j + P_l)$ or $P_j \le \gamma_j^{\mathrm{th}}P_l$, we can obtain \eqref{eq:out2,Kj=M}. The proof is completed.}

\section*{Appendix~C: Proof of Theorem \ref{theorem3}} \label{Appendix:C}
\renewcommand{\theequation}{C.\arabic{equation}}
\setcounter{equation}{0}

\textit{The receiver-specific outage probability of $U_l$ can be expressed as}
\begin{align}
    P_{U_l} = \underbrace{1 - \Pr \left( \gamma_{l \to i} \ge \gamma_i^{\mathrm{th}}, \gamma_{l \to j} \ge \gamma_j^{\mathrm{th}}, \gamma_l \ge \gamma_l^{\mathrm{th}} \right)}_{J_{3}}.
    \label{eq:C1}
\end{align}

\textit{By substituting \eqref{ljiancei}, \eqref{ljiancej}, and \eqref{ljiancezishen} into \eqref{eq:C1}, and letting $g_l = |h_l^{\mathrm{MRC}}|^2$, the expression for $J_{3}$ is formulated as}
\begin{equation}
    J_{3} = 1 - \Pr \left( g_l \ge \theta_1, g_l \ge \theta_2, g_l \ge \theta_3 \right),
\label{eq:C2}
\end{equation}
\textit{where $\theta_1 = \frac{\gamma_i^{\mathrm{th}}\sigma^2}{P_i - \gamma_i^{\mathrm{th}}(P_j + P_l)}$, $\theta_2 = \frac{\gamma_j^{\mathrm{th}}\sigma^2}{P_j - \gamma_j^{\mathrm{th}}P_l}$, and $\theta_3 = \frac{\gamma_l^{\mathrm{th}}\sigma^2}{P_l}$. Let $\xi_l = \max \{ \theta_1, \theta_2, \theta_3 \}$ for $P_i > \gamma_i^{\mathrm{th}}(P_j + P_l)$ and $P_j > \gamma_j^{\mathrm{th}}P_l$, $J_3$ can be calculated as}
\begin{align}
    J_{3} &= \Pr(g_l < \xi_l) \notag \\
    &= F_{g_l}(\xi_l).
\end{align}

\textit{If either $P_i \le \gamma_i^{\mathrm{th}}(P_j + P_l)$ or $P_j \le \gamma_j^{\mathrm{th}}P_l$, at least one required SIC step is infeasible for any finite $g_l$, and thus $P_{U_l} = 1$.}

\textit{Under the symmetric two-branch Nakagami-$\mu$ model, the MRC-combined channel power gain satisfies $g_l \sim \Gamma(2\mu, \Lambda_l)$ with $\Lambda_l = \frac{\mu}{\chi\Omega_l}$. Therefore, for the integer-valued Nakagami parameter adopted in the analysis, $J_{3}$ can be calculated using the CDF of $g_l$ as}
\begin{align}
    J_{3} = 1 - e^{-\Lambda_l \xi_l} \sum_{n=0}^{2\mu-1} \frac{(\Lambda_l \xi_l)^n}{n!}.
    \label{eq:C3}
\end{align}

\textit{Substituting \eqref{eq:C3} into \eqref{eq:C1}, together with the infeasible cases $P_i \le \gamma_i^{\mathrm{th}}(P_j + P_l)$ or $P_j \le \gamma_j^{\mathrm{th}}P_l$, we can obtain \eqref{eq:out3,Kl=M}. The proof is completed.}

\bibliographystyle{IEEEtran}
\bibliography{mybib}

\end{document}